\documentclass[11pt,leqno]{article}

\newcommand{\hugeDebug}{false}

\usepackage{epsfig}

\usepackage{xspace}
\usepackage{ifthen}

\usepackage{amssymb}

\usepackage{amsmath}

\usepackage{pifont}

\let\ensuremathTEMP\ensuremath

\makeatletter%
\def\nottoobig#1{{\hbox{$\left#1\vcenter to1.111\ht\strutbox{}\right.\n@space$}}}
\makeatother%

\makeatother%

\makeatletter

\newcount\hour  \newcount\minutes  \hour=\time  \divide\hour by 60
\minutes=\hour  \multiply\minutes by -60  \advance\minutes by \time
\def\mmmddyyyy{\ifcase\month\or Jan\or Feb\or Mar\or Apr\or May\or Jun\or Jul\or
  Aug\or Sep\or Oct\or Nov\or Dec\fi \space\number\day, \number\year}
\def\hhmm{\ifnum\hour<10 0\fi\number\hour :%
  \ifnum\minutes<10 0\fi\number\minutes}
\def\Draft{{\it Draft of \mmmddyyyy}}

\def\ps@jtsheadings{%
\def\@oddhead{\it\rightmark\hfil\rm\thepage}%
\def\@oddfoot{\hfil\Draft}%
\if@twoside
\def\@evenhead{\rm\thepage\hfil\it\leftmark}%
\def\@evenfoot{\Draft\hfil}%
\else
\let\@evenhead\@oddhead%
\let\@evenfoot\@oddfoot%
\fi
}
\def\ps@jtsplain{%
\def\@oddhead{\hfil\Draft}%
\def\@oddfoot{\hfil\rm\thepage\hfil}%
\let\@evenfoot\@oddfoot%
\if@twoside \def\@evenhead{\Draft\hfil} \else \let\@evenhead\@oddhead \fi
}

\def\chaptermark#1{\markboth{\thechapter.\ #1}{\thechapter.\ #1}}%
\def\sectionmark#1{\markright{\thesection.\ #1}}

\makeatother

\makeatletter \@beginparpenalty=10000 \makeatother
\def\underl#1 {\leavevmode\let\first=\relax\underli #1 }
\def\underli#1 {\ifx&#1\let\next=\relax\unskip
                \else\let\next=\underli\first\ulinebox{#1}\fi\let\first=\undersp\next}
\def\undersp{\penalty50\ulinebox{\space}\penalty50}
\def\ulinebox#1{\vtop{\hbox{\strut#1}\hrule}}
\def\unice#1 {\underl #1 & }

\def\desclabel#1{\bf #1\hfil}
\def\desc{\list{}{%
\setlength{\leftmargin}{0pt}
\labelwidth= \leftmargin
\advance \labelwidth by -\labelsep
\let \makelabel=\desclabel}}

\def\descHACKlabel#1{\bf #1\hfil}
\def\descHACK{\list{}{%
\setlength{\leftmargin}{0pt}
\labelwidth= \leftmargin
\advance \labelwidth by -\labelsep
\let \makelabel=\descHACKlabel}}

\newcounter{extremeleftlistcounter}
  {\begin{list}{\arabic{extremeleftlistcounter}~~~}{\usecounter{extremeleftlistcounter}%
        \setlength{\labelsep}{0pt}\setlength{\leftmargin}{0pt}%
        \setlength{\labelwidth}{0pt}\setlength{\listparindent}{0pt}}}%
  {\end{list}}

\newcounter{leftlistcounter}
  {\begin{list}{\arabic{leftlistcounter}~~~}{\usecounter{leftlistcounter}%
        \setlength{\labelsep}{0pt}\setlength{\leftmargin}{15pt}%
        \setlength{\labelwidth}{15pt}\setlength{\listparindent}{0pt}}}%
  {\end{list}}



\makeatletter 

\newcommand{\sbr}[2]{$   \left.\begin{array}{c}\vspace*{#2in}\end{array}\right\} #1 $}

\newlength{\filength}
\newsavebox{\gcbox}
\sbox{\gcbox}{\framebox[\filength]{\rule{0ex}{2ex}}}

\newlength{\leftjustindent}
\newlength{\@leftjustindent}
\def\leftjust{\let\\\@leftjustcr\let\end\@endleftjust
  \addtolength{\@leftjustindent}{\leftjustindent} \vcenter\bgroup
\halign\bgroup \hbox to\displaywidth{
\rule{\@leftjustindent}{0ex}$\displaystyle##$\hfill }\crcr }
\def\endleftjust{\crcr\egroup\egroup\endgroup}
\def\@endleftjust#1{\crcr\egroup\egroup\@checkend{#1}\endgroup}
\def\@leftjustcr{\crcr}

\newtheorem{theorem}{Theorem}[section]
\newtheorem{corollary}[theorem]{Corollary}

\newcommand{\qedblob}{\mbox{\rule[-1.5pt]{5pt}{10.5pt}}}
\renewcommand\qedblob{\ding{113}}

\def\literalqed{{\ \nolinebreak\hfill\mbox{\qedblob\quad}}}

\def\qed{\literalqed}

\newtheorem{lemma}[theorem]{Lemma}

\newtheorem{proposition}[theorem]{Proposition}
\newcommand{\singlespacing}{\let\CS=
\@currsize\renewcommand{\baselinestretch}{1}\tiny\CS}
\newcommand{\singlespacingplus}{\let\CS=
\@currsize\renewcommand{\baselinestretch}{1.25}\tiny\CS}
\newcommand{\doublespacing}{\let\CS=
\@currsize\renewcommand{\baselinestretch}{1.75}\tiny\CS}
\newcommand{\extradoublespacing}{\let\CS=
\@currsize\renewcommand{\baselinestretch}{1.9}\tiny\CS}
\newcommand{\hugedraftspacing}{\let\CS=
\@currsize\renewcommand{\baselinestretch}{2.4}\tiny\CS}
\makeatother

\ifthenelse{\equal{\hugeDebug}{false}}{
\newcommand{\normalspacing}{\niceonespacing}
}
{
\newcommand{\normalspacing}{\niceninespacing}
}

\mathcode`\0="0030      
\mathcode`\1="0031
\mathcode`\2="0032
\mathcode`\3="0033
\mathcode`\4="0034
\mathcode`\5="0035
\mathcode`\6="0036
\mathcode`\7="0037
\mathcode`\8="0038
\mathcode`\9="0039

\newtheorem{definition}[theorem]{Definition}

\makeatletter
\clubpenalty=\@highpenalty
\widowpenalty=\@highpenalty
\makeatother

\makeatletter
\newcommand{\niceonespacing}{\let\CS=\@currsize\renewcommand{\baselinestretch}{1.1}\tiny\CS}
\newcommand{\nicetwospacing}{\let\CS=\@currsize\renewcommand{\baselinestretch}{1.2}\tiny\CS}
\newcommand{\nicethreespacing}{\let\CS=\@currsize\renewcommand{\baselinestretch}{1.3}\tiny\CS}
\newcommand{\singlespacingplusplus}{\let\CS=\@currsize\renewcommand{\baselinestretch}{1.35}\tiny\CS}
\newcommand{\nicefourspacing}{\let\CS=\@currsize\renewcommand{\baselinestretch}{1.4}\tiny\CS}
\newcommand{\nicefivespacing}{\let\CS=\@currsize\renewcommand{\baselinestretch}{1.5}\tiny\CS}
\newcommand{\nicesixspacing}{\let\CS=\@currsize\renewcommand{\baselinestretch}{1.6}\tiny\CS}
\newcommand{\nicesevenspacing}{\let\CS=\@currsize\renewcommand{\baselinestretch}{1.7}\tiny\CS}
\newcommand{\niceeightspacing}{\let\CS=\@currsize\renewcommand{\baselinestretch}{1.8}\tiny\CS}
\newcommand{\niceninespacing}{\let\CS=\@currsize\renewcommand{\baselinestretch}{1.9}\tiny\CS}
\makeatother

\makeatletter
\def\@cite#1#2{[#1\if@tempswa , #2\fi]}
\makeatother

\makeatletter
\def\@citex[#1]#2{\if@filesw\immediate\write\@auxout{\string\citation{#2}}\fi
  \def\@citea{}\@cite{\@for\@citeb:=#2\do
    {\@citea\def\@citea{,\linebreak[0]}\@ifundefined
       {b@\@citeb}{{\bf ?}\@warning
       {Citation `\@citeb' on page \thepage \space undefined}}%
\hbox{\csname b@\@citeb\endcsname}}}{#1}}
\makeatother

\makeatletter
\def\@cite#1#2{[#1\if@tempswa , #2\fi]}
\def\@citex[#1]#2{\if@filesw\immediate\write\@auxout{\string\citation{#2}}\fi
  \def\@citea{}\@cite{\@for\@citeb:=#2\do
    {\@citea\def\@citea{,\kern1pt\linebreak[0]}\@ifundefined
       {b@\@citeb}{{\bf ?}\@warning
       {Citation `\@citeb' on page \thepage \space undefined}}%
\hbox{\csname b@\@citeb\endcsname}}}{#1}}
\makeatother

\makeatletter
\def\ps@thesis{\def\@oddhead{\hfil\rm\thepage\hfil}\def\@oddfoot{}\def\@evenhead{\hfil\rm\thepage\hfil}\def\@evenfoot{}\def\chaptermark##1{}\def\sectionmark##1{}}
\makeatother

\makeatletter
\def\foobarpt{\textfont\z@\tenrm 
  \scriptfont\z@\ninrm \scriptscriptfont\z@\sevrm
\textfont\@ne\tenmi \scriptfont\@ne\ninmi \scriptscriptfont\@ne\sevmi
\textfont\tw@\tensy \scriptfont\tw@\ninsy \scriptscriptfont\tw@\sevsy
\textfont\thr@@\tenex \scriptfont\thr@@\tenex \scriptscriptfont\thr@@\tenex
\def\unboldmath{\everymath{}\everydisplay{}\@nomath\unboldmath
          \textfont\@ne\tenmi 
          \textfont\tw@\tensy \textfont\lyfam\tenly
          \@boldfalse}\@boldfalse
\def\boldmath{\@ifundefined{tenmib}{\global\font\tenmib\@mbi\@magscale1\global
        \font\tensyb\@mbsy \@magscale1\global\font
         \tenlyb\@lasyb\@magscale1\relax\@addfontinfo\@xiipt
              {\def\boldmath{\everymath
                {\mit}\everydisplay{\mit}\@prtct\@nomathbold
                \textfont\@ne\tenmib \textfont\tw@\tensyb 
                \textfont\lyfam\tenlyb\@prtct\@boldtrue}}}{}\@xiipt\boldmath}%
\def\prm{\fam\z@\tenrm}%
\def\pit{\fam\itfam\tenit}\textfont\itfam\tenit \scriptfont\itfam\ninit
   \scriptscriptfont\itfam\sevit
\def\psl{\fam\slfam\tensl}\textfont\slfam\tensl 
     \scriptfont\slfam\tensl \scriptscriptfont\slfam\tensl
\def\pbf{\fam\bffam\tenbf}\textfont\bffam\tenbf 
   \scriptfont\bffam\ninbf \scriptscriptfont\bffam\ninbf 
\def\ptt{\fam\ttfam\tentt}\textfont\ttfam\tentt
   \scriptfont\ttfam\nintt \scriptscriptfont\ttfam\nintt 
\def\psf{\fam\sffam\tensf}\textfont\sffam\tensf
    \scriptfont\sffam\tensf \scriptscriptfont\sffam\tensf
\def\psc{\@getfont\psc\scfam\@xiipt{\@mcsc\@magscale1}}%
\def\ly{\fam\lyfam\tenly}\textfont\lyfam\tenly 
   \scriptfont\lyfam\ninly \scriptscriptfont\lyfam\sevly
 \@setstrut \rm}

\makeatother

\newcommand{\naturalnumber}{\ensuremath{{  \mathbb{N} }}}

\newcommand{\sharpp}{{\rm \#P}}

\newcommand{\up}{{\rm UP}}

\newcommand{\coup}{{\rm coUP}}

\newcommand{\p}{{\rm P}}

\newcommand{\clsp}{{\rm CL\#P}}
\newcommand{\clusp}{{\rm CLU\#P}}
\newcommand{\cluspcircular}{{\rm CLU\#P_{\rm circular}}}
\newcommand{\cluspcirculartilde}{{\widetilde{\rm CLU\#P}_{\rm circular}}}
\newcommand{\cluspfree}{{\rm CLU\#P_{\rm free}}}
\newcommand{\bottom}{{b}}
\renewcommand{\top}{{t}}
\newcommand{\lexprec}{{\prec_{\rm lex}}}
\newcommand{\notlexprec}{{\not\prec_{\rm lex}}}

\newcommand{\uprec}{{{\prec}}}
\newcommand{\lexleq}{{\leq_{\rm lex}}}
\newcommand{\leqlex}{{\lexleq}}
\newcommand{\zop}{{\{0,1\}^{p(|x|)}}}
\newcommand{\pox}{{{p(|x|)}}}
\newcommand{\zof}{{{\rm 0\hbox{-}1\hbox{-}F}}}
\newcommand{\domain}{{\mbox{\it{}domain}}}
\newcommand{\prefix}{{\mbox{\it{}prefix}}}
\newcommand{\suffix}{{\mbox{\it{}suffix}}}
\newcommand{\greatest}{{\mbox{\it{}greatest}}}
\newcommand{\least}{{\mbox{\it{}least}}}

\newcommand{\startofproof}{\noindent{\bf Proof}\quad}
\newcommand{\sproof}{\noindent{\bf Proof}\quad}
\newcommand{\startofproofof}[1]{\noindent{\bf Proof of {#1}}\quad}
\newcommand{\sproofof}[1]{\noindent{\bf Proof of {#1}}\quad}

\newcommand{\pair}[1]{\mathopen\langle{#1}\mathclose\rangle}

\newcommand{\sigmastar}{\ensuremath{\Sigma^\ast}}

\newcommand{\fpt}{\ensuremath{\fp_{\rm t}}}

\newcommand{\bigo}{{\protect\cal O}}

\newcommand{\condition}{\,\ensuremathTEMP{\mbox{\large$|$}}\:}

\def\land{{\; \wedge \;}}

\newenvironment{block}{\begin{list}{\hbox{}}{\leftmargin 1em
    \itemindent -1em \topsep 0pt \itemsep 0pt \partopsep 0pt}}{\end{list}}

\ifthenelse{\equal{\hugeDebug}{false}}{
\setlength{\oddsidemargin}{0.25in}
\setlength{\evensidemargin}{\oddsidemargin}
\setlength{\textwidth}{6in}
\setlength{\textheight}{8in}
\setlength{\topmargin}{-0.0in}
}
{
\setlength{\oddsidemargin}{-0.4in}
\setlength{\evensidemargin}{\oddsidemargin}
\setlength{\textwidth}{5.3in}
}

\ifthenelse{\equal{\hugeDebug}{false}}{
\pagestyle{plain}
}
{
\pagestyle{headings}
}

\makeatletter
\def\@listI{\leftmargin\leftmargini \parsep 4.5pt plus 1pt minus 1pt\topsep
6pt plus 2pt minus 2pt \itemsep  2pt plus 2pt minus 1pt}

\let\@listi\@listI
\@listi
\makeatother

\newcommand{\upsvp}{\ensuremath{{\rm UPSV_p}}}
\newcommand{\upsvt}{\ensuremath{{\rm UPSV_t}}}
\newcommand{\upsv}{\ensuremath{{\rm UPSV}}}

\newcommand{\npsvt}{\ensuremath{{\rm NPSV_t}}}
\newcommand{\npsvp}{\ensuremath{{\rm NPSV_p}}}

\makeatletter 
 \newcommand{\setoffdisplay}{\rule{5.9in}{1pt}}

\makeatother



\newcommand{\acc}{\ensuremathTEMP{{  acc}}}
\newcommand{\numacc}{\ensuremathTEMP{{ \# acc}}}

\newcommand{\fp}{\ensuremathTEMP{{\rm FP}}}

\title{
Cluster Computing and the Power of Edge Recognition
}

\author{
\emph{Lane A. Hemaspaandra}\protect\thanks{Supported
in part by grant
NSF-CCF-0426761.
Work done in part while 
visiting
  Julius-Maximilians-Universit\"at 
W\"{u}rzburg.
URL: \mbox{\tt{}www.cs.rochester.edu/u/lane}.
}\\
  Department of Computer Science\\University of Rochester\\
             Rochester, NY 14627
\\USA
\and 
\emph{Christopher M. Homan}\thanks{URL: \mbox{\tt{}www.cs.rit.edu/$\tilde{~}$cmh}.}\\Department of Computer Science\\
Rochester Institute of Technology\\Rochester, NY 14623
\\ USA
\and 
\emph{Sven
Kosub}\thanks{URL: {\tt{}www14.in.tum.de/personen/kosub}.}\\ Institut f\"ur
Informatik\\ Technische Universit\"at M\"unchen\\ 
D-85748 Garching b.~M\"unchen
\\ Germany
}

\date{
September 16, 2005}

\begin{document}
\normalspacing
\typeout{WARNING:  BADNESS used to suppress reporting.  Beware.}
\hbadness=3000
\vbadness=10000 
\bibliographystyle{alpha}
{\singlespacing\maketitle}

\vfill

%
\begin{abstract}
  {\singlespacing We study the robustness---the invariance under
    definition changes---of the cluster class
    $\clsp$~\cite{hem-hom-kos-wag:tr-special-with-backpointer-for-cluster-new-paper-to-icalp-version:interval-functions}.
    This class contains each $\sharpp$ function that is computed by a
    balanced Turing machine whose accepting paths always form a
    cluster with respect to some length-respecting total order with
    efficient adjacency checks.  The definition of $\clsp$ is heavily
    influenced by the defining paper's focus on (global) orders.  In
    contrast, we define a cluster class, $\clusp$, to capture what
    seems to us a more natural model of cluster computing.  We prove
    that the naturalness is costless: $\clsp = \clusp$.  Then we
    exploit the more natural, flexible features of $\clusp$ to prove
    new robustness results for $\clsp$ and to expand what is known
    about the closure properties of $\clsp$.

    The complexity of recognizing edges---of an ordered collection of
    computation paths or of a cluster of accepting computation
    paths---is central to this study.  Most particularly,
    our proofs exploit the power of unique discovery of edges---the
    ability of nondeterministic functions to, in certain settings,
    discover on exactly one (in some cases, on at most one)
    computation path a critical piece of information regarding edges
    of orderings or clusters.
    
  }
\end{abstract}

\vfill



\section{Introduction}
Cluster computing, in the complexity-theoretic use of the term, was
introduced by Kosub 
in~\cite{kos:j:clusters}, though
he
notes there that there was earlier work that
focused in a rather different sense on cluster-like behavior
(\cite{wag:j:bounded,vol-wag:j:opt} and we mention in passing that the
so-called telescoping normal form of the boolean
hierarchy~\cite{cai-gun-har-hem-sew-wag-wec:j:bh1} and the parallel
census technique of Selman~\cite{sel:t:adaptive,gla-hem:j:clarityI}
also provide early examples of the type of behavior Kosub was there
observing, namely, settings in which ``yes'' answers always occur in a
contiguous block).  In particular, Kosub defined and studied the
class $\rm c\sharpp$, which is the set of all $\sharpp$ functions
computed by (i.e., given by the number of accepting paths of)
lexicographical cluster machines---loosely put,
machines such that on each input, all
the accepting paths are lexicographically adjacent to each other (they
form a contiguous block).  He obtained quite comprehensive results, but
they depended critically on the very simple structure of lexicographical
order, namely, that if one knows the left and right edges of a
lexicographical cluster, it is easy to compute the size of the cluster.

Yet the underlying motivating issue---to what extent does requiring
that all accepting paths be closely related in some order restrict the
ability of nondeterministic Turing machines to compute $\sharpp$
functions?---certainly is not tied to the artificial simplicity of
lexicographical order.  Just as
self-reducibility~\cite{sch:c:self-reducible,mey-pat:t:int} has
not only
been defined with respect to a focus on the lexicographical order and
decreasing chains with respect to length there (as
in~\cite{bal-dia-gab:b:sctI-2nd-ed,bal-dia-gab:b:sctII}) but also (and
most elegantly) has been defined with respect to having polynomially
length-bounded decreasing chains within appropriate more general
classes of orders (as in \cite{mey-pat:t:int,sch:c:self-reducible}),
%
so also is it natural to study cluster computing with respect to more
flexible ordering.

To imagine how to naturally do this, we think of 
the model underlying $\rm c\sharpp$, which, again, is the 
class of
functions that are the
numbers of accepting computation paths of 
balanced (a Turing machine is 
balanced if there is some polynomial $p$ such that on
each input $x$ it holds that each nondeterministic path has exactly
$p(|x|)$ binary nondeterministic guesses)
Turing machines in which the
accepting paths are always lexicographically adjacent.
So the accepting block on
a given input is, assuming any paths accept, just a lexicographically
contiguous block among the length $p(|x|)$ strings, where one
views---as we do throughout this paper---each accepting path (on a
given input) as being named by its nondeterministic guesses.
Intuitively speaking, we suggest that it might be very natural to
generalize this by keeping essentially the entire setting mentioned
above, except on input $x$ viewing the strings at length $p(|x|)$ not
as being in lexicographical order, but rather viewing them as follows.
For each balanced nondeterministic machine whose number of accepting
paths defines a function in our new class, there must be
polynomial-time 
computable
functions $\bottom$ (the bottom function), $\top$ (the
top function), and $\prec$ (the adjacency function) such that: We view
$\bottom(x) \in \{0,1\}^{p(|x|)}$ as the least string of length
$p(|x|)$; $\uprec(x,y,z)$ tells whether on input $x$ the string $z \in
\{0,1\}^{p(|x|)}$ comes immediately after $y \in \{0,1\}^{p(|x|)}$
in our linear ordering of the length $p(|x|)$ strings; if one using
those two functions starts at $b(x)$ and moves through one string
after another under the adjacency rule specified by
$\uprec(x,\cdot,\cdot)$, one goes though each string of length
$p(|x|)$ and ends up at $\top(x)$; and if there are any accepting
paths on input $x$, then all the accepting paths on input $x$ form a
cluster---a contiguous block---within this ordering.  In particular,
regarding the ordering, we allow an arbitrary linear ordering of the
length $p(|x|)$ strings subject to it being easy to tell the biggest
and smallest elements in our new order, and to recognize adjacency in
our new order.  Let us call the class thus defined $\clusp$.

Though we suggest that the $\clusp$ definition and model are very easy
to work with, it is very important to note that a previous paper
already defined a generalization of Kosub's notion with exactly the
goal of handling more general orderings.  In particular, this was done
by~\cite{hem-hom-kos-wag:tr-special-with-backpointer-for-cluster-new-paper-to-icalp-version:interval-functions},
resulting in the class $\clsp$.  $\clsp$'s definition, however, is
heavily influenced by the overall focus of that paper on global orders
(rather than input-specific orderings).  In particular, that paper
requires all inputs to have their computation paths share the
\emph{same} order with respect to defining what it means to be a
cluster.  For example, if on input $x$ computation paths $y$ and $z$
exist and $y \prec z$ (respectively, $y \not\prec z$), then for each
input $x'$ on which those computation paths exist (namely, all strings
$x'$ on which the nondeterminism polynomial happens to evaluate to the
same value on $|x'|$ as it does on $|x|$, and so certainly for all
strings $x'$ of the same length as $x$) it must also hold that $y
\prec z$ (respectively, $y\not\prec z$).  Further, the fact that that
paper really requires a global---over all of $\sigmastar$---order
forces the ordering for each input $x$ to smoothly link the strings
related to computation on input $x$ to the other, utterly irrelevant
paths.  Although these constraints are arguably reasonable in a paper
whose focus is on global, total orders (in the formal sense), we here
suggest that if one were to simply take the idea of Kosub and
shuffle\footnote{Throughout this paper, we use ``shuffle'' in its
  common-language sense of permuting a single collection, rather than
  in the very different sense in which the term is sometimes used in
  theoretical computer science, namely, taking two separate lists and
  interleaving them.}  the paths that apply to that input, the notion
of $\clusp$ would seem 
a more natural approach to and model of doing that.

Fortunately, one does not have to choose between the classes $\clsp$
and $\clusp$.  This is because our main result is that the new class
$\clusp$, which was defined to directly capture a natural, local,
machine-directed notion of cluster computing, has exactly the same
descriptive power as the class $\clsp$, which is based on a global
shared order: $\clsp = \clusp$.  This result is in
Section~\ref{s:main-equality}, which also shows another robustness
result that will be central to our later study of two other
notions---free cluster and circular cluster machines.  That other
robustness result is essentially that unambiguity of cluster edge
recognition is sufficient to ensure that even seemingly more flexible
models in fact generate just the $\clsp$ functions.

Section~\ref{s:closures}, partially by using our newfound freedom to
study $\clsp$ by studying $\clusp$, shows a number of closure
properties of $\clsp$.  For example,
\cite{hem-hom-kos-wag:tr-special-with-backpointer-for-cluster-new-paper-to-icalp-version:interval-functions}
proved that if $\clsp$ is closed under increment then $\up=\coup$, and
we show that the converse holds.

Our model, $\clusp$-type machines, has $\fpt$ (total, polynomial-time
computable) top and bottom elements on each input.
Section~\ref{s:free} studies two alternate models.  $\cluspfree$
removes any explicit complexity requirement regarding the top and
bottom elements.  $\cluspcircular$ requires the ordering on our
computation paths to be circular---thus there is no top or bottom
element.  We prove a number of results about these classes, and most
particularly we show 
(a)~that $\up=\coup$ is a sufficient condition for 
$\clsp = \cluspfree =
\cluspcircular$,
and 
(b)~that $\up=\coup$ is a necessary condition for
$\clsp = \cluspfree$, and even for 
$\cluspfree \subseteq \cluspcircular$.  Result~(b) can be viewed 
as reasonably strong evidence that 
$\cluspfree$ is a strictly more powerful, flexible class than
$\clsp$, and can also be viewed as reasonably strong 
evidence that 
some $\cluspfree$ functions are not in $\cluspcircular$.
So freeing the endpoints from their 
$\fpt$ constraint seems to yield a real increase in 
descriptive power.

The proofs in this paper are thematically linked.  Most of them focus
on the power of what we will call ``unique discovery'' of facts about
about top and bottom elements and about greatest and least accepting
paths---i.e., about ``edges.''  By unique discovery we mean that
critical pieces of edge-related information used in our proofs are
partial or total $\upsv$ (unambiguous polynomial-time single-valued)
functions~\cite{gro-sel:j:complexity-measures,kos:j:clusters}.
Informally speaking, we mean that our proof strategy will often be:
\begin{enumerate}
\item\label{step:first-step} Seek to guess some critical piece of
  information (such as the right edge of a cluster).
\item If we succeeded on the current path in guessing that information
  correctly, do FOO and otherwise do BAR,
\end{enumerate}
and, critically, our settings will variously ensure that in
step~(\ref{step:first-step}) either exactly one or at most one path
guesses the critical information, that that path ``knows'' it has done
so (i.e., could write on an output tape the information and set a bit
declaring it has successfully obtained the information), and each
other path knows that it has not done so.

\section{Definitions}\label{s:definitions}
$\Sigma=\{0,1\}$ will be our alphabet.  The boolean relation
$\lexprec$ is defined as: $a \lexprec b$ is true when $b$ is the
lexicographical successor of $a$ and is false otherwise, e.g.,
$111\lexprec 0000$ and $010 \lexprec 011$, but $00 \notlexprec 11$.
We use NPTM as a shorthand for ``nondeterministic polynomial-time
Turing machine.''  As is common, for a given nondeterministic machine
$M$ and a string $x$, $\acc_M(x)$ denotes the set of accepting paths
of machine $M$ on input $x$, and $\numacc_M(x)$ is defined as
$||\acc_M(x)||$.  $\fpt$ denotes the total, polynomial-time computable
functions (usually from $\sigmastar$ to $\sigmastar$).

Given any string $x \in \sigmastar$ and any integer $n \leq |x|$,
$\prefix(x,n)$ denotes the first $n$ bits of $x$ and $\suffix(x,n)$
denotes the last $n$ bits of $x$.  If $n > |x|$ these functions are
undefined.

For each polynomial $p$ and each NPTM $M$, $M$ will be said to be
$p$-balanced (see~\cite{kos:j:clusters}) exactly if for each input $x$
the set of nondeterministic guesses along the computation paths of $M$
is precisely $\{0,1\}^{p(|x|)}$.  That is, $M$ on input $x$ has
exactly $2^{p(|x|)}$ computation paths, one corresponding to each
possible guess of $p(|x|)$ bits.  Note that we do not require that
each step of the machine involves a nondeterministic guess.


We turn immediately to defining the central class of this paper,
$\clusp$.  In defining $\clusp$, we seek to keep Kosub's notion of a
cluster as a block of adjacent paths, but we allow that adjacency to
be with respect to a ``shuffling'' of the paths, rather than to have
to be with respect to lexicographical order.  However, the shuffle must
be simple enough that in polynomial time we can get the first and last
paths' names, and also in polynomial time we can, given paths $q$ and
$r$, determine whether the path immediately greater than (i.e.,
right-adjacent to) $q$ is $r$.  And a function $f$ belongs to $\clusp$
if the function gives the number of accepting paths of an
(appropriately balanced) Turing machine whose accepting paths always
form a cluster of this sort.  Although the formal definition is a bit
intimidating, we stress that it is merely rigorously capturing this
intuitively simple notion.

\begin{definition}\label{d:clusp}
  A (total) function $f: \sigmastar \rightarrow \naturalnumber$
  belongs to $\clusp$ if $(\exists \mbox{ polynomial }p) \allowbreak
  (\exists \mbox{ $p$-balanced NPTM }M) \allowbreak ( \exists
  \,\,\bottom,\top \in \fpt) \allowbreak (\exists$ 3-argument,
  polynomial-time computable predicate $\prec) \allowbreak ( \forall
  x) \allowbreak (\exists \mbox{ bijection }h_x \mbox{ from }
  \Sigma^{p(|x|)} \mbox{ to } \Sigma^{p(|x|)} )$ such that:
\begin{enumerate}
\item $|\bottom(x)| = |\top(x)| = p(|x|)$.
  
\item $h(\bottom(x)) = 0^{p(|x|)} \land h(\top(x)) =
  1^{p(|x|)}$.
  
\item $(\forall y,z \in \{0,1\}^{p(|x|)}) [\uprec(x,y,z)
  \iff h_x(y)\lexprec h_x(z)]$.
  
\item All accepting paths are clustered with respect to
  $\uprec(x,\cdot,\cdot)$.
  That is, if $f(x) \neq 0$
  then $(\exists \ell,u \in \{0,1\}^{p(|x|)}) [\acc_{M}(x)
  = \{w \in \zop\condition \allowbreak h_x(\ell)\leqlex h_x(w) \lexleq
  h_x(u)\}]$.
  
\item $f(x) = \numacc_M(x)$.
\end{enumerate}

\end{definition}

As mentioned in the introduction, even for two same-length strings $x$
and $y$, it is completely possible that $\uprec(x,\cdot,\cdot)$ and
$\uprec(y,\cdot,\cdot)$ will differ dramatically.  That is, $\clusp$
focuses heavily on reordering the paths related to the given input,
and just those paths, and indeed may do so in a way that can vary
based on the input.  (Though formally speaking the definition above
requires $\prec$ to be defined on all input triples, it is easy to see
from the above definition that on input $x$ all that matters is what
$\uprec(x,\cdot,\cdot)$ does when its second two arguments are
distinct strings in $\zop$.  For all other inputs, we can typically
just ignore $\prec$'s output or view it as being false.)

We now turn to the definition of
$\clsp$~\cite{hem-hom-kos-wag:tr-special-with-backpointer-for-cluster-new-paper-to-icalp-version:interval-functions}.
That definition requires the entire universe of paths---over all
inputs to a machine---to be embedded in a single, shared order.  As
noted earlier, this limits one in two ways: 
the obvious constraint
that one must embed paths over different inputs into the same order
(and so when inputs have the same length, their paths must be
identically shuffled) and a more subtle side-constraint that even
though all computation paths of a machine on a given input are of the
same length, in this setting the adjacency test must work even between
that length and other lengths, i.e., all of $\sigmastar$ must be woven
into a single, giant order with the right feasibility properties.

To support the definition of $\clsp$, we briefly define some related
notions
(see~\cite{hem-hom-kos-wag:tr-special-with-backpointer-for-cluster-new-paper-to-icalp-version:interval-functions},
from which we take these definitions essentially word for word, for
consistency), namely, ``length-respecting total order $A$'' and
``$A$-cluster.''  A binary relation $A \subseteq \Sigma^* \times
\Sigma^*$ is a {partial order} if it is reflexive, antisymmetric
(i.e., $(\forall x,y \in\Sigma^*)[ x \neq y \implies ((x,y)\not\in A
\lor (y,x) \not\in A)]$), and transitive.  A partial order $A$ is a
{total order} if, for all $x,y \in \Sigma^*$, $(x,y) \in A$ or $(y,x)
\in A$.
We write $x\prec_A y$ if $x<_A y$ and there is no $z$ such that $x<_A
z<_A y$. If $x\prec_A y$, we say that $x$ {is left-adjacent to} $y$
or, equivalently, $y$ {is right-adjacent to} $x$.
Let $M$ be NPTM that is $p$-balanced for some polynomial $p$.  Let $y$
and $z$ encode computation paths of $M$ on $x$.  By the above
assumption that $M$ is balanced, $|y|=|z|$. Fix a total order $A$ on
$\Sigma^*$. We say that $y\sim_{A,M,x} z$ if and only if (a) $y\prec_A
z$ or $z\prec_A y$, and (b) $M$ on $x$ accepts on path $y$ if and only
if $M$ on $x$ accepts on path $z$.  Let $\equiv_{A,M,x}$ be the
equivalence closure (i.e., the reflexive-symmetric-transitive closure)
of $\sim_{A,M,x}$. Then the relation $\equiv_{A,M,x}$ is an
equivalence relation and thus induces a partitioning of the
computation tree of $M$ on $x$. An $A$-cluster is an equivalence class
whose representatives are accepting paths.
Additionally, we consider $\emptyset$ to be a valid
$A$-cluster.
An order $A$ on $\Sigma^*$ is said to be
length-respecting if and only if, for all $x,y$, $|x|<|y|$ implies
$x<_A y$.

\begin{definition}[\cite{hem-hom-kos-wag:tr-special-with-backpointer-for-cluster-new-paper-to-icalp-version:interval-functions}]\label{d:clsp} 
  A function $f$ belongs to the class $\clsp$ if there exist a
  polynomial $p$, a $p$-balanced NPTM $M$, and a length-respecting
  total order $A$ with efficient adjacency checks such that, for all
  $x$, the following conditions hold:
\begin{enumerate}
\item The set of all accepting paths of $M$ on $x$ is an $A$-cluster.
\item $f(x)=\numacc_M(x)$.
\end{enumerate}
\end{definition}

We now define the classes $\cluspfree$ and $\cluspcircular$.  Their
definitions are similar to that of $\clusp$.  However, $\cluspfree$
removes the constraint that top- and bottom-finding must be
polynomial-time computable, though $\uprec$ will implicitly create top
and bottom elements.  $\cluspcircular$ makes the order be a circular
order, thus removing any notion of ``top'' and ``bottom.''

\begin{definition}\label{d:cluspfree}
  A (total) function $f: \sigmastar \rightarrow \naturalnumber$
  belongs to $\cluspfree$ if $(\exists \mbox{ polynomial }p)
  \allowbreak (\exists \mbox{ $p$-balanced NPTM }M) \allowbreak
  (\exists$ 3-argument, polynomial-time computable predicate $\prec)
  \allowbreak ( \forall x) \allowbreak (\exists \mbox{ bijection }h_x
  \mbox{ from } \Sigma^{p(|x|)} \mbox{ to } \Sigma^{p(|x|)} )$ such
  that:
\begin{enumerate}

  
\item $(\forall y,z \in \{0,1\}^{p(|x|)}) [\uprec(x,y,z)
  \iff h_x(y)\lexprec h_x(z)]$.
  
\item All accepting paths are clustered with respect to
  $\uprec(x,\cdot,\cdot)$.
  That is, if $f(x) \neq 0$
  then $(\exists \ell,u \in \{0,1\}^{p(|x|)}) [\acc_{M}(x)
  = \{w \in \zop\condition \allowbreak h_x(\ell)\leqlex h_x(w) \lexleq
  h_x(u)\}]$.
  
\item $f(x) = \numacc_M(x)$.
\end{enumerate}

\end{definition}

\begin{definition}\label{d:circular}
  A (total) function $f: \sigmastar \rightarrow \naturalnumber$
  belongs to $\cluspcircular$ if $(\exists \mbox{ polynomial }p)
  \allowbreak (\exists \mbox{ $p$-balanced NPTM }M) \allowbreak
  (\exists$ 3-argument, polynomial-time computable predicate $\prec)
  \allowbreak ( \forall x) \allowbreak (\exists \mbox{ bijection }h_x
  \mbox{ from } \Sigma^{p(|x|)} \mbox{ to } \Sigma^{p(|x|)} )$ such
  that:
\begin{enumerate}
\item\label{c:circularity} $(\forall y,z \in \{0,1\}^{p(|x|)})$
  $$[\uprec(x,y,z) \iff (h_x(y)\lexprec h_x(z) \lor (h_x(y)=1^{p(|x|)}
  \land h_x(z)=0^{p(|x|)}))].$$
  
\item\label{p:diverge} All accepting paths are clustered with respect to
  $\uprec(x,\cdot,\cdot)$.
  That is, if $f(x) \neq 0$
  then $(\exists \ell,u \in \{0,1\}^{p(|x|)}) [\acc_{M}(x)
  = \{w \in \zop\condition \allowbreak h_x(\ell)\leqlex h_x(w) \lexleq
  h_x(u)\}]$.
  
\item $f(x) = \numacc_M(x)$.
\end{enumerate}
\end{definition}

The reader may reasonably
worry that our definition of $\cluspcircular$ is cheating.
In particular, one may worry that 
Definition~\ref{d:circular}'s
part~\ref{c:circularity} has the adjacency definition go ``around the
corner'' (that is, it adjacency-links
$0^{p(|x|)}$
and $1^{p(|x|)}$ in the 
under-the-image-of-$h$ space),
but that 
Definition~\ref{d:circular}'s
part~\ref{p:diverge}
doesn't similarly allow the accepting paths to go
``around the corner,'' and that this is a somewhat strange and striking
asymmetry of approach between those two aspects of the definition.
However,
note that in the definition of a $\cluspcircular$ function we can
without loss of generality
require that the preimage of the bijection $h_x$ has the property that
$h_x^{-1}(0^{p(|x|)})$ is an accepting path of the machine (on that 
input) if any
accepting paths exist (on that input). 
That is, the first condition in Definition~\ref{d:circular} 
is invariant under cyclic shifts of the numbering
$h_x$ of the elements in $\{0,1\}^{p(|x|)}$.
So the above-mentioned  worry about the
definition turns out, upon some thought, not to be a worry at all.
Indeed, later in the paper this observation will be a useful
feature, namely, in the proof of Proposition~\ref{p:complementary}.

Note that it follows immediately from the definitions that $\clusp\subseteq
\cluspfree$ and $\clusp\subseteq \cluspcircular$.

Finally, let us state the definitions of the function classes $\upsvt$
and $\upsvp$~\cite{gro-sel:j:complexity-measures,kos:j:clusters},
which are the (respectively total and partial) unambiguous versions of
the central, single-valued nondeterministic function classes $\npsvt$
and
$\npsvp$~\cite{boo-lon-sel:j:quant,boo-lon-sel:j:qual,sel:j:taxonomy}.
When speaking of nondeterministic machines as computing (possibly
partial) functions from $\sigmastar$ to $\sigmastar$, we view each
path as having no output if the path is a rejecting path, and if a
path is an accepting path then it is viewed as outputting whatever
string $s\in\sigmastar$ is on the output tape (along that path) when
that path halts.  A (potentially partial) function $f:\sigmastar
\rightarrow \sigmastar$ belongs to $\upsvp$ if there is an NPTM $M$
that (a)~on each input has at most one accepting path, (b)~on each
input $x$ on which $M$ has exactly one accepting path, $f(x)$ is the
output on that path, and (c)~on each input $x$ on which $M$ has no
accepting paths, $f(x)$ is undefined (i.e., $\domain(f)=\{x\condition
M(x)$ has no accepting paths$\}$).  A function $f:\sigmastar
\rightarrow \sigmastar$ belongs to $\upsvt$ if $f$ belongs to $\upsvp$
and $f$ is total.  $\upsvp$ and $\upsvt$ functions capture the flavor
of ``unique discovery,'' and will (often implicitly and sometimes
explicitly) be central in our proofs.

\section{Robustness of CLU\#P}\label{s:main-equality}
In this section, we study the robustness of $\clusp$.  $\clusp$ on its
surface might seem to be far more flexible than $\clsp$, given that
unlike $\clsp$ it is not chained by the requirement of a global order
and the related need to have same-length strings' paths coexist in the
same order and to link consistently between lengths.\footnote{One
  might note that, on the other hand, $\clsp$ lacks the $\fpt$
  constraints (on the top and bottom elements among the computation
  paths) that $\clusp$ obeys, and in that way at least potentially
  might seem to have some flexibility that $\clusp$ might lack.
  However, though $\clsp$ does not explicitly speak of top and bottom
  functions at each length, it is not hard to see that it has top and
  bottom functions (mapping from each $x$---or even from
  $0^{|x|}$---to the top and bottom elements at length $\pox$) that
  are computable in $\upsvt$.  We will show later in this section that
  $\clusp$ remains unchanged if one allows its top and bottom
  functions to be drawn not just from $\fpt$ but even from $\upsvt$.
  Thus, $\clusp$ is not at a disadvantage on this issue.}
Nonetheless, we now prove that these two classes are equal: $\clsp =
\clusp$.

Briefly put, to show that $\clusp \subseteq \clsp$ we tie together the
exponential number of orderings (over all inputs sharing the same path
length).  To show that $\clsp \subseteq \clusp$, we uniquely discover
the top and the bottom elements and then embed into a broader search
space a clone of the action of our $\clsp$ machine on the current
input.

\begin{theorem}\label{t:two}$\clusp = \clsp$.
\end{theorem}

We first prove a simple lemma.  We do so in part because it will be
helpful in the proof (though one could work around it if needed), and
mostly because the proof provides a
simple initial
example of how to prove things about cluster classes.

We say a polynomial $p$ is monotonic exactly if, for all natural
numbers $n$, $p(n)<p(n+1)$.
\begin{lemma}\label{l:monotonic} If $f\in\clusp$, then $f\in\clusp$ via 
  some integer-coefficient nondeterminism polynomial (in the sense of
  Definition~\ref{d:clusp}) that is monotonic.
\end{lemma}

\startofproof Let $f\in\clusp$.  Let $p$, $M$, $\prec$, $\top$, and
$\bottom$ capture $f$ in the sense of Definition~\ref{d:clusp}.  If
$p$ is monotonic, then we are done.  Note that asymptotically the
polynomial $p$ cannot diverge to negative infinity and can never take
on a negative value.
So there will exist a monotonic, integer-coefficient polynomial $p'$
such that $(\forall n \in \naturalnumber) [ 0 \leq p(n) < p'(n)]$.
$f$ will be computed in the sense of Definition~\ref{d:clusp} by
$p'$-balanced NPTM $M'$, $\bottom'$, $\top'$, and $\uprec'$, each
defined as follows.  

\begin{figure}[!tbp+]
\begin{center}
  \input{figure-dual-monotonic.pstex_t}
\end{center}
\caption{\label{figure:easy}
  Figure for the proof of Lemma~\ref{l:monotonic}.  Key: R denotes
  paths that certainly are rejecting paths.}
\end{figure}

$\bottom'(x) = 0^{p'(|x|)-p(|x|)}\bottom(x)$ and
$\top'(x) = 1^{p'(|x|)}$.
NPTM $M'$ works as follows: On input $x$, $M'$ guesses
${p'(|x|)-p(|x|)}$ bits, call them $\alpha$, and then guesses $p(|x|)$
bits, call them $\beta$.  If $\alpha \in 0^*$ and $M(x)$ accepts along
computation path $\beta$ (recall that we speak of paths by naming
their nondeterministic guess bits), then we accept, and otherwise we
reject.
Define the predicate $\uprec'$ as follows: 
$\uprec'(x,y,z)$ will evaluate to true exactly if
$|y|=|z|=p'(|x|)$ and either
\begin{enumerate}
\item $\prefix(y, p'(|x|)-p(|x|)) = \prefix(z, p'(|x|)-p(|x|)) = 0^{
    p'(|x|)-p(|x|)}$ and $\uprec(x,\suffix(y,p(|x|)),\suffix(z,p(|x|))$,
  or
  
\item $\prefix(y, p'(|x|)-p(|x|)) \neq 0^{ p'(|x|)-p(|x|)}$ and $y
  \lexprec z$, or
  
\item $y = 0^{ p'(|x|)-p(|x|)} \top(x)$ and $z = 0^{
    p'(|x|)-p(|x|)-1}1 0^{p(|x|)}$.
\end{enumerate}

That is, we guess dummy bits, simulate the underlying machine $M$ in
the leftmost subtree, and weave all the paths naturally together by
inheriting the adjacency operator for the leftmost subtree, and for
the rest we follow lexicographical order (and are careful at the
boundary about the connection between the leftmost subtree and the
rest).

Figure~\ref{figure:easy}a shows this proof pictorially, with
paths shown (left to right) in lexicographical order.
Figure~\ref{figure:easy}b pictures the same construction, but in the
way we will use from now on in all our figures.  Namely,
Figure~\ref{figure:easy}b shows paths, left to right, not in
lexicographical order, but rather ``pre-permuted'' into our order.  (So
in part~b of the figure, the ``inherited order'' is a guide to 
how the figure has been pre-permuted.  But that just is an issue of
our illustration.  Far more critical is to keep in mind that what
really is inherited in that segment is the order itself---which paths
should be considered adjacent to which.  Also, to be clear, 
the curved arrow on the bottom of 
Figure~\ref{figure:easy}a 
is denoting a single 
adjacency---that the path $0^{p'(|x|)-p(|x|)}t(x)$ is left-adjacent
to the path
$0^{p'(|x|)-p(|x|)-1}10^{p(|x|)}$.  But the curved arrow on
the bottom of 
Figure~\ref{figure:easy}b 
indicates that in the pre-permuted picture shown the adjacencies 
sweep one at a time from
$0^{p'(|x|)-p(|x|)}b(x)$ 
up to 
$0^{p'(|x|)-p(|x|)}t(x)$ in the order inherited from the 
underlying order we are building upon.)~\qed

\smallskip

\sproofof{Theorem~\ref{t:two}}
For $\clusp \subseteq \clsp$, let $f \in \clusp$. Let $p$, $M$, $\prec$, 
$\top$, and $\bottom$ capture $f$ in the sense of Definition~\ref{d:clusp}. 
By Lemma~\ref{l:monotonic}, we may assume without loss of generality that 
$p$ is monotonic. We show that $f \in \clsp$ by constructing $p'$, $M'$, and 
length-respecting total order $A'$ that capture $f$ in the sense of 
Definition~\ref{d:clsp}. Define $p'$ on input 
$n \in \mathbb{N}$ as $p(n) + n$.

\begin{figure}[!tbp+]
\begin{center}
  \input{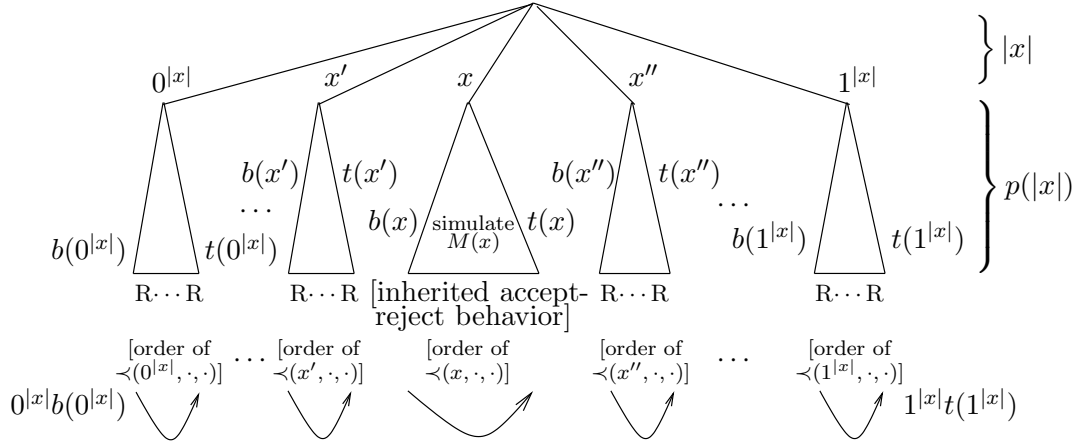}
\end{center}
\caption{\label{figure:clucl}
  Figure for the $\clusp \subseteq \clsp$ part of the
  proof of Theorem~\ref{t:two}.  Key: R denotes
  paths that certainly are rejecting paths and $x'$ (respectively,
$x''$) is the string lexicographically preceding (respectively,
succeeding) $x$.}
\end{figure}

$M'$ works as follows: $M'$ on input $x\in\Sigma^*$ nondeterministically 
guesses $|x|$ bits $y$. If $y = x$ then
$M'$ simulates $M$ on input $x$, where each nondeterministic branch accepts 
iff the corresponding branch in $M$ accepts. If $y \neq x$ then $M'$ 
nondeterministically guesses $p(|x|)$ bits but then ignores them and rejects. 
Clearly, each nondeterministic branch of $M'$ uses exactly $p'(|x|)$ guess
bits. 

$A'$ is the same as the lexicographical 
order, except for those strings that, for some $n\in\mathbb{N}$, are of length
$p'(n)$. For those strings, $0^n\bottom(0^n)$ comes first and $1^n\top(1^n)$ 
comes last. For all $x,y \in \Sigma^{n + p(n)}$,
$x \prec_A y$ iff 
\begin{enumerate}
\item $\prefix(x,n) = \prefix(y,n)$ and $\uprec(\prefix(x,n),\suffix(x,p(n)),
\suffix(y,p(n)))$, or
\item $\prefix(x,n) \lexprec \prefix(y,n)$ and $\suffix(x,p(n)) = 
\top(\prefix(x,n))$ and $\suffix(y,p(n)) = \bottom(\prefix(y,n))$.
\end{enumerate}
Figure~\ref{figure:clucl} shows pictorially this part of the proof.

For $\clsp \subseteq \clusp$, let $f \in \clsp$. Let $p$, $M$, and $A$ 
capture $f$ in the sense of Definition~\ref{d:clsp}.

\begin{figure}[!tbp+]
\begin{center}
  \input{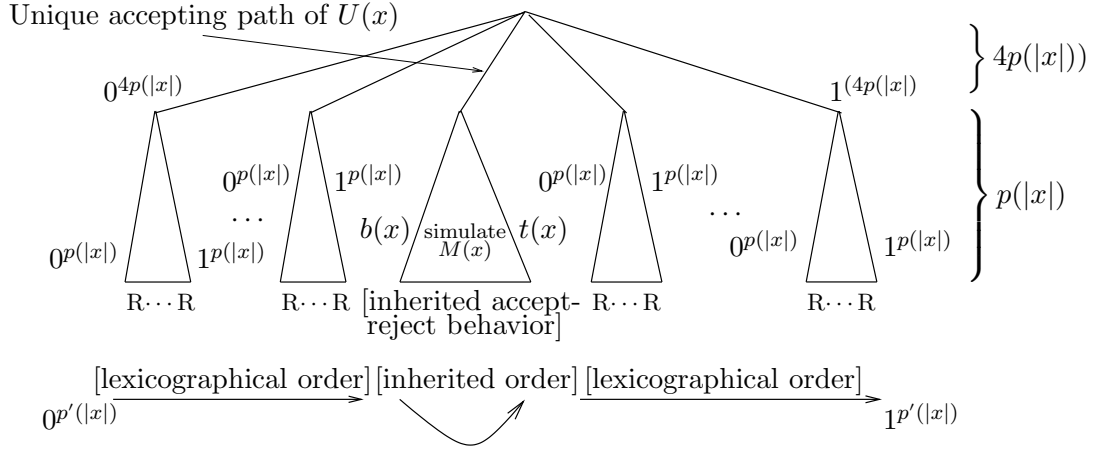}
\end{center}
\caption{\label{figure:clclu}
  Figure for 
$\clsp \subseteq \clusp$ part of 
the proof of Theorem~\ref{t:two}.  Key: R denotes
  paths that certainly are rejecting paths.}
\end{figure}

Let $U$ be an NPTM
that works as follows: On input $x \in \Sigma^*$, $U$ nondeterministically 
guesses strings $u,y,z \in\Sigma^*$, where $|y| = |z| = p(|x|)$ and 
$|u| = 2p(|x|)$. If $p(|x|) = 0$ then accept. If $p(|x|) > 0$, 
$\prefix(u,\pox-1)\prec_A y$, and $z \prec_A \suffix(u,\pox+1)$ then accept. 
Otherwise reject. 
Clearly, $U$ on any input has exactly one accepting path (recall that 
$A$ is length-respecting), is balanced for polynomial $4p$, 
and we can in polynomial time determine, for any 
$x\in\Sigma^*$ and $y \in \acc_U(x)$, what the least and greatest (with respect
to $A$) strings of length $p(|x|)$ are.

We show
$f \in \clusp$ by providing $p'$, $M'$, $\prec'$, $\top'$, and $\bottom'$,
defined below,
that capture $f$ in the sense of Definition~\ref{d:clusp}. Let $p'= 5p$. For
any $x \in \Sigma^*$,
let $\bottom'(x) = 0^{p'(|x|)}$ and $\top'(x) = 1^{p'(|x|)}$. 

$M'$ works as follows: On input $x \in \Sigma^*$, $M'$ nondeterministically
guesses $w,y \in\Sigma^*$, where $|w| = 4p(|x|)$ and $|y| = p(|x|)$. If
$w \in \acc_U(x)$  and $y\in\acc_M(x)$, then accept. 
Otherwise, reject.

Predicate $\prec'$ is defined as follows. For $x,u,u' \in \Sigma^*$,
$\uprec'(x,u,u')$ evaluates to true exactly if, for some $v,w,y,z \in \Sigma^*$,
where $|v| = |w| = 4p(|x|)$ and $|y| = |z| = p(|x|)$, $u = vy$, $u' = wz$,
it holds that:
\begin{enumerate}
\item $\{v,w\} \cap \acc_U(x) = \emptyset$ and $vy \lexprec wz$, or
\item $\{v,w\} \subseteq \acc_U(x)$ ($\Rightarrow v=w$) and $y\prec_A z$, or
\item $v \lexprec w$ and $w \in \acc_U(x)$ and $y = 1^{p(|x|)}$ and $z$ is the 
minimum (with respect to $A$) of the set of all computational paths of $M$ on 
$x$, or
\item $v \lexprec w$ and $v\in \acc_U(x)$ and $z = 0^{p(|x|)}$ and $y$ is the 
maximum (with respect to $A$) of the set of all computational paths of $M$ on 
$x$.
\end{enumerate}
Figure~\ref{figure:clclu} shows pictorially this part of the proof.~\qed

\smallskip

We now derive a robustness result that might seem a bit less natural
than Theorem~\ref{t:two}.  However, this robustness result provides a
critical tool for proving natural results and gives substantial
insight into what suffices to make cluster computation simple.

To state the result, we must define notions of the greatest element
and the least element of an accepting path cluster.  The slight
unnaturalness occurs in the circular model, in particular in the case
when all paths are accepting paths since in that case, even though
there is no natural choice of greatest and least accepting paths, our
definition makes a choice.

Note that we are speaking not about top and bottom notions among all
paths of a given length, but rather are seeking the greatest and least
\emph{accepting} paths with respect to a given input and the ordering
implicit in $\uprec$.

Let $p$, $M$, and $\prec$ be a nondeterminism polynomial, machine, and
adjacency predicate in either the $\cluspfree$ model or the
$\cluspcircular$ model.  We define two partial functions ($p$ is
implicit in $M$, but for uniformity and clarity in settings like this
we include $p$ throughout the paper) $\greatest_{p,M,\uprec}$ and
$\least_{p,M,\uprec}$ as follows.  Let $M$ compute the function $f$,
i.e., on input $x$, $f(x) = \numacc_M(x)$.

If $f(x)=0$, then (in both the free and the circular models)
$\greatest_{p,M,\uprec}(x)$ and $\least_{p,M,\uprec}(x)$ are
undefined.  In the free model, if $f(x) \neq 0$, then
$\greatest_{p,M,\uprec}(x)$ is the unique length $p(|x|)$ string $z$
that is an accepting path to which no length $p(|x|)$ accepting path
is right-adjacent
(i.e., the unique string $z$ of length $p(|x|)$ such that $z$ is an
accepting path of $M$ on input $x$ and yet $(\forall w \in \zop)[w \in
\acc_M(x) \implies \neg \uprec(x,z,w)]$).  Similarly, in the free
model, if $f(x) \neq 0$, then $\least_{p,M,\uprec}(x)$ is the unique
length $p(|x|)$ string $z$ that is an accepting path to which no
length $p(|x|)$ accepting path is left-adjacent.

The above definitions will not work in the circular model if $M$, on
input $x$, accepts on all paths, as there the definition would give
nothing, but for our proofs we cannot allow that to happen.  It
actually is fine to break this impasse by saying that when that
happens, the greatest function takes on the value of any of $M$'s
accepting paths such that the least function has as its value the path
right-adjacent to that path.  However, for clarity and specificity, we
sacrifice a bit of flexibility and choose one particular
impasse-breaking splitting point as follows for the circular model.
If $f(x) = 0$, $\greatest_{p,M,\uprec}(x)$ and
$\least_{p,M,\uprec}(x)$ still are undefined.  If $f(x)=2^{\pox}$ then
$\greatest_{p,M,\uprec}(x) = 1^\pox$.  If $f(x)=2^{\pox}$ and $\pox =
0$ then $\least_{p,M,\uprec}(x) = \epsilon$.
If $f(x)=2^{\pox}$ and $\pox
\neq 0$ then $\least_{p,M,\uprec}(x)$ equals the unique length $\pox$
string $z$ satisfying $\uprec(x,1^\pox,z)$.  In the circular model, if
$0\leq f(x) < 2^\pox$, $\greatest$ and $\least$ are defined exactly as
in the free model.

\begin{theorem}\label{t:edge-power}
\begin{enumerate}
\item\label{t:edge-power-free} Let $f$ be computed by $p$, $M$, $\uprec$ in the free model.  If
  $\greatest_{p,M,\uprec}\in\upsvp$ and
  $\least_{p,M,\uprec}\in\upsvp$, then $f\in\clusp$.
\item\label{t:edge-power-circ} Let $f$ be computed by $p$, $M$, $\uprec$ in the circular model.  If
  $\greatest_{p,M,\uprec}\in\upsvp$ and
  $\least_{p,M,\uprec}\in\upsvp$, then $f\in\clusp$.
\end{enumerate}
\end{theorem}

That is, unique discovery of boundaries is sufficient in both the free
and the circular models to remove any power beyond that of $\clusp$.

Briefly summarized, our proof will seek to uniquely discover the
greatest and least accepting paths, and on the (at most one) block
that discovers them will simulate the original machine except with
each path sheathed in three dummy rejecting paths.
Blocks that fail to make the unique discovery will follow
lexicographical order, and the unique discovery block (which will exist
exactly when $f(x)>0$) will adopt a somewhat complex order that allows
us to indeed be $\clusp$-like.

\startofproofof{Theorem~\ref{t:edge-power}}
The proofs for items~\ref{t:edge-power-free} and \ref{t:edge-power-circ} are
essentially the same.
Let $f$ be computed by $p$, $M$, and $\uprec$ in either the free 
or 
the 
circular
 model.
Suppose for polynomials $q$ and $q'$ that $\greatest_{p,M,\uprec}\in\upsvp$ 
via $q$-balanced NPTM $U$ and
 $\least_{p,M,\uprec}\in\upsvp$ via $q'$-balanced NPTM $U'$. We now define
$p'$, $M'$, $\uprec'$, $\bottom'$, and $\top'$ that capture $f$ in the sense of 
Definition~\ref{d:clusp}.  
Let $p'(n) = q(n) + q'(n) + p(n) + 2$. 

$M'$ works as follows. On input $x \in \Sigma^*$, $M'$ nondeterministically
guesses strings $y,z,u,v \in \Sigma^*$, where $|y| = q(|x|)$, $|z| = q'(|x|)$, 
$|u| = p(|x|)$, and 
$|v| = 2$. If $y \in \acc_U(x)$, $z\in\acc_{U'}(x)$, $u\in\acc_M(x)$, and 
$v = 01$ then accept. Otherwise reject.

\begin{figure}[!tbp+]
\begin{center}
  \input{figure-edge-power-broad.pstex_t}
\end{center}
\caption{\label{figure:edge-power-broad}
  First figure for the proof of Theorem~\ref{t:edge-power}.  Key: R denotes
  paths that certainly are rejecting paths. The string $y$ is the accepting 
path of $U(x)$ concatenated with the accepting path of $U'(x)$, if such paths
 exist. The string $y'$ (respectively,
$y''$) is the string lexicographically preceding (respectively,
succeeding) $y$.}
\end{figure}
\begin{figure}[!tbp+]
\begin{center}
  \input{figure-edge-power-detailed.pstex_t}
\end{center}
\caption{\label{figure:edge-power-detailed}
  Second figure for the proof of Theorem~\ref{t:edge-power}.  Key: R denotes
  paths that certainly are rejecting paths and A denotes accepting
paths. String $y$ is as in Figure~\ref{figure:edge-power-broad}.}
\end{figure}

We define $\bottom'$ on input $x \in \Sigma^*$ as
\[\bottom'(x) = \left\{\begin{array}{ll}0^{q(|x|)+q'(|x|)}\least_{p,M,\uprec}(x)01&
\mbox{if $0^{q(|x|)} \in \acc_U(x)$ and $0^{q'(|x|)} \in \acc_{U'}(x)$}\\
0^{p'(|x|)}&\mbox{otherwise.}\end{array}\right.\]
On input $x \in \Sigma^*$, $\top'$ is defined as $1^{p'(|x|)}$.

We define $\prec'$, on inputs $x,w,w' \in \Sigma^*$ as follows. Predicate 
$\uprec'(x,w,w')$ evaluates to true exactly if there exist strings 
$y, y', u, u', v, v' \in \Sigma^*$ such that $|y| = |y'| = q(|x|)+q'(|x|)$, 
$|u| = |u'| = p(|x|)$, $|v| = |v'| = 2$,
$w = yuv$, $w' = y'u'v'$, and 
it holds that:
\begin{enumerate}
\item $\{w,w'\} \cap \acc_{M'}(x) = \emptyset$ and $w \lexprec w'$, or
\item $\{w,w'\} \subseteq \acc_{M'}(x)$ and $\uprec(x,u,u')$, or
\item $w \in \acc_{M'}(x)$ and $u=\greatest_{p,M,\prec}(x)$ and 
$\prefix(w,q(|x|)+q'(|x|)) = \prefix(w',q(|x|)+q'(|x|))$ and 
$\suffix(w',p(n)+2) \in 0^*$, or
\item $(\exists w'' \in \acc_{M'}(x))[w\lexprec w''\lexprec w')]$, or
\item $w' \in \acc_{M'}(x)$ and $u' \in \least_{p,M,\prec}(x)$ and 
$y \lexprec y'$ and $\suffix(w,p(n)+2) \in 1^{\pox + 2}$.
\end{enumerate}
Figures~\ref{figure:edge-power-broad} and \ref{figure:edge-power-detailed} 
show how this construction works.~\qed

\smallskip

We will employ Theorem~\ref{t:edge-power} in Section~\ref{s:free}, but
let us note now that focusing on the boundaries of the accepting block
is enough to speak to issues regarding the complexity of the top and
bottom functions.

\begin{corollary}\label{c:UPSV-flexible}
  If in Definition~\ref{d:clusp} ``$\exists \,\,\bottom,\top \in
  \fpt$'' is replaced with ``$\exists \,\,\bottom,\top \in \upsvt$,''
  the class defined by the new definition remains precisely $\clusp$.
\end{corollary}

\sproof Let $p$, $M$, $\uprec$, $\bottom$, and $\top$ satisfy the
definitions of $\clusp$ except altered as noted in the statement of
this corollary.

Notice that on a given input $x$, a computation path $\rho$ of length
$\pox$ is the value of $\least_{p,M,\uprec}$ if $\rho$ is an accepting
path and either (a)~$x$ is in the domain of (the $\upsvt$ function)
$\bottom$ and $\bottom(x)=\rho$, or
(b)~there is a rejecting computation path that is left-adjacent to
$\rho$ (and note that if such a path exists it is unique).  Thus,
keeping in mind that if there are no accepting paths at length $\pox$
on input length the test just described will not select any path as
$\least_{p,M,\uprec}$, it is clear that by guessing each length $\pox$
path, testing that it is accepting, and then doing the above test and
outputting the path if the test succeeds, we have shown that
$\least_{p,M,\uprec}\in\upsvp$.  By a similar argument,
$\greatest_{p,M,\uprec}\in\upsvp$.  So by Theorem~\ref{t:edge-power}
the function computed by $p$, $M$, and $\uprec$ is in $\clusp$.~\qed

\section{Closure Properties of CLU\#P}\label{s:closures}
Arithmetic closure properties are not the focus of this paper.
However, in this section we briefly study some as an example of the
power of unique discovery of boundaries and to take advantage of the
fact that Theorem~\ref{t:two} allows us to prove closure properties of
$\clsp$ via the easier to work with model of $\clusp$.  In particular,
we show that an implication
of~\cite{hem-hom-kos-wag:tr-special-with-backpointer-for-cluster-new-paper-to-icalp-version:interval-functions}
is in fact a complete characterization.
\begin{theorem}[\cite{hem-hom-kos-wag:tr-special-with-backpointer-for-cluster-new-paper-to-icalp-version:interval-functions}]\label{t:clsp-up=coup}
  If $\clsp$ (equivalently in light of Theorem~\ref{t:two}, $\clusp$)
  is closed under increment (i.e., $f \in \clsp \implies
  (\lambda x. f(x)+1) \in \clsp$), then $\up=\coup$.
\end{theorem}

We prove that the converse holds and in fact prove that $\up=\coup$
characterizes a number of closures of $\clusp$.  We say a function is
natural-number-valued if it maps from $\sigmastar$ to
$\naturalnumber$.  All $\clusp$ functions are natural-number-valued.

\begin{theorem}\label{t:closures}
  The following statements are equivalent:
\begin{enumerate}
  
\item\label{it:closed:up=coup} $\up=\coup$.
  
\item\label{it:closed:incr} $\clusp$ is closed under increment.
  
\item\label{it:closed:fpt} $\clusp$ is closed under addition of 
natural-number-valued
  $\fpt$ functions.
  
\item\label{it:closed:upsvt} $\clusp$ is closed under addition of 
natural-number-valued
  $\upsvt$ functions.
  
\item\label{it:closed:addition} $\clusp$ is closed under addition.

\end{enumerate}
\end{theorem}

\sproof Clearly, all natural-number-valued $\upsvt$ functions
are $\clusp$ functions, so~\ref{it:closed:addition} $\Rightarrow$ 
\ref{it:closed:upsvt}. All $\fpt$ functions are $\upsvt$ so 
\ref{it:closed:upsvt} $\Rightarrow$~\ref{it:closed:fpt}. Clearly,
\ref{it:closed:fpt} $\Rightarrow$~\ref{it:closed:incr}. By 
Theorem~\ref{t:clsp-up=coup},~\ref{it:closed:incr} $\Rightarrow$
\ref{it:closed:up=coup}. 

To prove~\ref{it:closed:up=coup} 
$\Rightarrow$~\ref{it:closed:addition}, suppose $\up=\coup$.
Let $f \in \clusp$ via $M_f$, $p_f$, $\bottom_f$, $\top_f$ and $\prec_f$
and let $g \in \clusp$ via  $M_g$, $p_g$, $\bottom_g$, $\top_g$ and $\prec_g$.
Let $i \in \{f,g\}$. Let $V_i$ be an NPTM that, on input $x \in \Sigma^*$,
nondeterministically guesses $y,z,u,v \in \Sigma^*$, where 
$|y|=|z|=|u|=|v|=p_i(|x|)$, and accepts (on the current path) 
if and only if 
$\{z,u\}\subseteq\acc_{M_i}(x)$ and both of
the following hold:
\begin{enumerate}
\item $(y \not\in\acc_{M_i}(x) \wedge \uprec_i(x,y,z))$ or 
$(z = \bottom_i(x) \wedge y \in 0^*)$, and
\item  $(v \not\in\acc_{M_i}(x) \wedge \uprec_i(x,u,v))$ or 
$(u = \top_i(x) \wedge v \in 0^*)$.
\end{enumerate}
Clearly, $V_i$ has on any input at most one accepting path, and if $V_i$ accepts
on input $x$ then the string $z$ (respectively, $u$) guessed by the
accepting path is $\least_{p_i,M_i,\prec_i}(x)$ (respectively, 
$\greatest_{p_i,M_i,\prec_i}(x)$). Assuming $\up=\coup$, $\overline{L(V_i)} \in 
\up$. Suppose for some polynomial $q_i'$ that $V_i'$ 
is a $q_i'$-balanced NPTM that decides $\overline{L(V_i)}$ and has on any 
input at most one 
accepting path. Let $U_i$ be a Turing machine that, on input $x \in \Sigma^*$,
nondeterministically guesses a string $z$ of length $q_i(|x|)$, where $q_i$
is a polynomial such that 
$(\forall n \in \naturalnumber)[q_i(n) > 4p_i(n) + q_i'(n)]$, 
and accepts (on the current path) if and 
only if $(\exists y \in \Sigma^*)[(y \in \acc_{V_i}(x) \vee y \in 
\acc_{V_i'}(x)) \wedge z \in y0^*)]$. Note that on any input
$U_i$ has at most one accepting path and that we can in polynomial
time determine, for any $x\in\Sigma^*$ and $y \in \acc_{U_i}(x)$, whether 
$i(x) > 0$
and, if so, what $\greatest_{p_i,M_i,\prec_i}(x)$ and $\least_{p_i,M_i,\prec_i}(x)$
are.

Let $U$ be an NPTM that, on input $x\in\Sigma^*$, guesses $y \in \Sigma^*$ of 
length $q_f(|x|) + q_g(|x|)$ and accepts (on the current path) 
exactly if $(\prefix(y,q_f(|x|)) \in 
\acc_{U_f}(x) \wedge \suffix(y,q_g(|x|)) \in \acc_{U_g}(x))$. Clearly,
$U$ has on any input exactly one accepting path, and 
we can, in polynomial time, determine, for any $x \in \Sigma^*$,
$y \in \acc_U(x)$, and $i \in \{f,g\}$ whether $i(x) > 0$
and, if so, what $\least_{p_i,M_i,\prec_i}(x)$ and 
$\greatest_{p_i,M_i,\prec_i}(x)$ are.

We now construct $M$, $p$, $\bottom$, $\top$, and $\uprec$ that capture 
the function 
$f + g$
in the sense of Definition~\ref{d:clusp}. Let 
$p(n) = q_f(n) + q_g(n) + q(n) + 3$, 
where $q$ is a polynomial such that $(\forall n \in \naturalnumber)
[q(n) > p_f(n) + p_q(n)]$.

$M$ on input $x \in \Sigma^*$ works as follows: $M$ nondeterministically
guesses $y \in \Sigma^*$, where $|y| = p(|x|)$, and accepts (on the 
current path) exactly if
 $\prefix(y,q_f(|x|)+q_g(|x|)) \in \acc_U(x)$ and:
\begin{enumerate}
\item For some $z \in \acc_{M_f}(x)$, $\suffix(y,q(|x|) + 3) \in 0z0^*1$, 
or
\item for some $z \in \acc_{M_g}(x)$, $\suffix(y,q(|x|) + 3) \in 1z0^*1$.
\end{enumerate}
Clearly, 
$(\forall x)[\numacc_M(x) = f(x) + g(x)]$. Note that our accepting paths always have at 
least two ``extra'' bits at the end of the path. This allows us to sheath, in 
a manner similar
to what we did in the proof of Theorem~\ref{t:edge-power}, each accepting
path in at least three dummy rejecting paths.

Before defining $\prec$, we will first define $\greatest_{p,M,\prec}$ and 
$\least_{p,M,\prec}$. For $x \in \Sigma^*$, let $u \in \acc_U(x)$ and let
\[\greatest_{p,M,\prec}(x) = \left\{\begin{array}{ll}
u1\greatest_{p_g,M_g,\prec_g}(x)0^{q(|x|)-p_g(x)+1}1&\mbox{if 
$\greatest_{p_g,M_g,\prec_g}(x)$ is defined,}\\
u0\greatest_{p_f,M_f,\prec_f}(x)0^{q(|x|)-p_f(x)+1}1&\mbox{otherwise, if }
\greatest_{p_f,M_f,\prec_f}(x)\\
&\mbox{is defined,}\\
\mbox{undefined} & \mbox{otherwise.}\end{array}\right.\]
and let
\[\least_{p,M,\prec}(x) = \left\{\begin{array}{ll}
u0\least_{p_f,M_f,\prec_f}(x)0^{q(|x|)-p_f(x)+1}1&\mbox{if 
$\least_{p_f,M_f,\prec_f}(x)$ is defined,}\\
u1\least_{p_g,M_g,\prec_g}(x)0^{q(|x|)-p_g(x)+1}1&\mbox{otherwise, if }
\least_{p_g,M_g,\prec_g}(x) \mbox{ is }\\
&\mbox{defined,}\\
\mbox{undefined} & \mbox{otherwise.}\end{array}\right.\]

\begin{figure}[!tbp+]
\begin{center}
\hspace*{-1.1in}\input{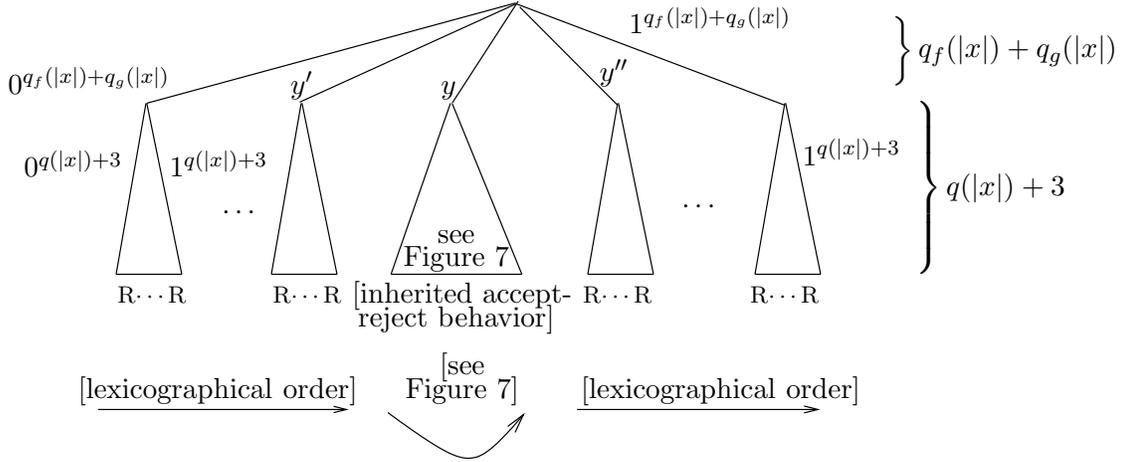}
\end{center}
\caption{\label{figure:closure-broad}
  First figure for the proof of Theorem~\ref{t:closures}.  Key: R denotes
  paths that certainly are rejecting paths. The string $y$ is the accepting 
path of $U(x)$. The string $y'$ (respectively,
$y''$) is the string lexicographically preceding (respectively,
succeeding) $y$.}
\end{figure}
\begin{figure}[!tbp+]
\begin{center}
  \input{figure-closure-detailed.pstex_t}
\end{center}
\caption{\label{figure:closure-detailed}
  Second figure for the proof of Theorem~\ref{t:closures}.  Key: R denotes
  paths that certainly are rejecting paths and A denotes accepting
paths. String $y$ is as in Figure~\ref{figure:closure-broad}.}
\end{figure} 

Define $\prec$ on $x, y, z \in \Sigma^*$ as follows (we claim
that our definition will result in $\greatest_{p,M,\prec}$ and
$\least_{p,M,\prec}$ as defined above): $\uprec(x,y,z)$ evaluates 
to true exactly if $|y| = |z| = \pox$ and:
\begin{enumerate}
\item $\{y,z\} \cap \acc_M(x) = \emptyset$ and $y\lexprec z$, or 
\item $\{y,z\} \subseteq \acc_M(x)$ and, for $w\in\acc_U(x)$:
  \begin{enumerate}
  \item for $u,v \in \acc_{M_f}$, $y \in w0u0^*1$ and $z \in w0v0^*1$ and 
$\uprec_f(x,u,v)$, or
    \item for $u,v \in \acc_{M_g}$, $y \in w1u0^*1$ and $z \in w1v0^*1$ and 
$\uprec_g(x,u,v)$, or
\item $f(x) > 0$ and $g(x) > 0$ and $y \in w0\greatest_{p_f,M_f,\prec_f}(x)0^*1$ 
and $z \in w1\least_{p_g,M_g,\prec_g}(x)0^*1$,
  \end{enumerate}
or
\item $\{y,z\} \cap \acc_M(x) = \emptyset$ and, for some $w \in \acc_M(x)$, 
$y\lexprec w \lexprec z$, or
\item $z \in \acc_M(x)$ and $z = \least_{p,M,\prec}(x)$ and $\prefix(y,q_f(|x|)+q_g(|x|)) \lexprec 
\prefix(z,q_f(|x|)+q_g(|x|))$ and $\suffix(y,q(|x|)+3) \in 1^*$, or
\item $y \in \acc_M(x)$ and $y = \greatest_{p,M,\prec}(x)$ and $\prefix(z,q_f(|x|)+q_g(|x|)) \in 
\acc_U(x)$ and $\suffix(z,q(|x|)+3) \in 0^*$.
\end{enumerate}

On input $x \in \Sigma^*$, $\bottom$ is defined as
\[\bottom(x) =\left\{
\begin{array}{ll}
\least_{p,M,\uprec}(x)&\mbox{if }0^{q_f(|x|)+q_g(|x|)}\in \acc_U(x)\\
&\mbox{and $\least_{p,M,\uprec}(x)$ is defined,}\\
0^{\pox}&\mbox{otherwise,}
\end{array}\right.
\]
and $\top(x)$ is defined as $1^{\pox}$.
Figures~\ref{figure:closure-broad} and \ref{figure:closure-detailed} 
show how this construction works.~\qed

\smallskip

Theorem~\ref{t:closures} may be viewed as evidence that 
$\clusp$ lacks various closure properties, e.g., closure 
under increment.  In contrast, the following result 
provides a closure property, proper decrement, that 
$\clusp$ possesses unconditionally.

\begin{theorem}\label{t:positive_closures}
$\clusp$ is closed under proper
decrement (i.e., $f\in\clusp \Longrightarrow 
(\lambda x.\max\{0,f(x)-1\})\in\clusp$).
\end{theorem}

\sproof Let $f\in\clusp$ via NPTM $M$, nondeterminism polynomial $p$, predicate $\prec$, and functions $b,t\in\fpt$. Define $U$ to be a machine that, on input $x\in\Sigma^*$, guesses a string
$y\in\Sigma^*$ of length $2p(|x|)$ and accepts (on the 
current path) if and only if $\prefix(y,p(|x|))\in \acc_M(x)$ and 
\begin{enumerate}
\item $\suffix(y,p(|x|))\notin\acc_M(x)$ and 
$\uprec(x,\prefix(y,p(|x|)),\suffix(y,p(|x|)))$, or
\item $\prefix(y,p(|x|))=t(x)$ and $\suffix(y,p(|x|))=0^{p(|x|)}$.
\end{enumerate}
Clearly, $U$ can have at most one accepting path. (Notice that $U$ can
be viewed as a balanced $\upsvp$ machine that computes
$\greatest_{p,M,\uprec}$.) Now consider an NPTM $N$ that, on input
$x\in\Sigma^*$, guesses a string $y$, where $|y|=3p(|x|)$, and accepts
(on the current path)
if and only if $\prefix(y,2p(|x|))\in \acc_U(x) \land
\suffix(y,p(|x|))\in \acc_M(x) \land
\prefix(y,p(|x|))\not=\suffix(y,p(|x|))$. It is clear that $N$ would
have as many accepting paths as $M$ has except regarding $N$'s analog
of the greatest accepting path of $M$ (if $M$ has accepting paths at
all), which is rejected by $N$. Thus, for all $x\in\Sigma^*$,
$\#\acc_N(x)=\max\{0,\#\acc_M(x)-1\}$. It remains to show how to
define an appropriate predicate $\prec'$: For $x,y,z\in\Sigma^*$,
$\uprec'(x,y,z)$ evaluates to true if and only if $|y|=|z|=3p(|x|)$
and
\begin{enumerate}
\item $\prefix(y,2p(|x|))=\prefix(z,2p(|x|))$ and $\uprec(x,\suffix(y,p(|x|)),\suffix(z,p(|x|)))$, or
\item $\prefix(y,2p(|x|))\lexprec \prefix(z,2p(|x|))$ and $\suffix(y,p(|x|))=t(x)$ and 
        $\suffix(z,p(|x|))=b(x)$.
\end{enumerate}
Furthermore, for all $x\in\Sigma^*$, define $b'(x)=0^{2p(|x|)}b(x)$ and $t'(x)=1^{2p(|x|)}t(x)$.
So NPTM $N$, nondeterminism polynomial $3p$, predicate $\prec'$, and $\fpt$ functions $b',t'$ witness the fact that $\#acc_N\in\clusp$.~\qed

\section{Free Cluster and Circular Cluster Computation}\label{s:free}
 
We defined $\cluspfree$ and $\cluspcircular$ in Section \ref{s:definitions}. Are these seemingly more flexible models truly more powerful than $\clusp$?
 We have not been able to prove that, though we will later, as
 Theorems~\ref{t:free-circ-eq}
 and~\ref{t:circ_eq_regular_consequence}, show that unless they
 are more powerful, certain collapses and closures hold.
 On the other hand,
 we now prove that $\up=\coup$ is sufficient to reduce the power of
 these two seemingly more flexible classes to that of $\clusp$.

\begin{theorem}\label{t:model-collapse}
  $\up=\coup\implies \clusp=\cluspfree=\cluspcircular$.
\end{theorem}

\sproof Let us first consider the $\cluspfree$ model.  Suppose we are
given some function $f$ in $\cluspfree$, along with $p$, $M$, and
$\uprec$ modeling it.  Note that (on input $x$) recognizing length
$\pox$ elements as being \emph{not} the top element is a UP
test.\footnote{\protect\label{footnote:relative}In this proof we
  always are speaking relative to $x$.  so this statement actually
  means that $\{\pair{x,y}\, \,| ~ 0\neq |y| = p(|x|) \land (\exists z
  \in \{0,1\}^{\pox})[\uprec(x,y,z)]\}$.}  Namely, if the length of
our paths is zero, then the path is the top element exactly if it is
$\epsilon$ (i.e., it is the path containing no guesses).  And
otherwise our arbitrary, given length $\pox$ path is not the top
element exactly if there exists a length $\pox$ element that is
right-adjacent to it with respect to $\uprec(x,\cdot,\cdot)$.
So being the top element is a coUP test (again, relative to $x$; see
Footnote~\ref{footnote:relative}).
And note that for each $x$ there is only one top element at length
$\pox$.
So, if $\up = \coup$, we have a $\upsvt$ function that finds the top
element on input $x$: Guess each length $\pox$ path, and then simulate
the $\up$ test for the $\coup$ question of whether it is the top
element.  Analogously, if $\up = \coup$, we have a $\upsvt$ function
that finds the bottom element.  By Corollary~\ref{c:UPSV-flexible} we
are done.

We now turn to the $\cluspcircular$ model.  Suppose we are given some
function $f$ in $\cluspcircular$, along with the $p$, $M$, $\uprec$
modeling it.  We ignore the case $\pox = 0$ as that is an
uninteresting and easy to handle special case.

Notice that it is a UP test to determine (on input $x$) whether there
exists \emph{any} right edge to the accepting paths of length $\pox$,
i.e., whether there exists any accepting path such that its (unique)
right-adjacent
path is a rejecting path.  So we now describe, under the assumption
that $\up=\coup$, a $\upsvp$ function computing
$\greatest$.  The lack of a right edge is a coUP test, and
$\up=\coup$.  And as just mentioned the existence of a right edge is a
UP test.

So our $\upsvt$-type machine guesses both that there is and that
there is not a right edge, and under each such guess it runs the
appropriate UP test.  If there is a right edge, some single path finds
that that is the case (and it is easy to ensure that that path in fact
even has the name of the right edge, as that is \emph{why} that test
is in UP in the first place).  Output the name of our right edge.  If
there is no right edge, again some single path discovers that that is
the case.  Since $\pox \neq 0$ and the path has found that there is no
right edge, and our order is circular, there are only two subcases:
All length $\pox$ paths reject or all length $\pox$ paths accept.  So
on the current path our machine does the following to distinguish
these two subcases.  Our machine tests whether $0^\pox$ is an
accepting path.  If $0^\pox$ is not an accepting path, then there are
no accepting paths at the current length, and so our machine rejects
on the current path.  If $0^\pox$ is an accepting path, then there are
only accepting paths at the current length, and so our machine in
order to obey the definition of $\greatest$ outputs $1^\pox$.

We have shown that in the circular case the partial function
$\greatest$ is in $\upsvp$ if $\up=\coup$.  The argument that in the
circular case the partial function $\least$ is in $\upsvp$ if
$\up=\coup$ is quite similar, except it focuses on the left edge of
the accepting block.  Also, it has one minor twist.  In the final
subcase ($\pox \neq 0$, we have found that there is no left edge, and
$0^\pox$ is an accepting path), we must output the unique length
$\pox$ string that is right-adjacent to $1^\pox$.  However, doing this
still within a $\upsvp$ function is easy, since there indeed is
(recall, $\pox \neq 0$) exactly one such string and so we can guess
and recognize it.

So, under the assumption $\up=\coup$, in the circular model
$\greatest$ and $\least$ are both $\upsvp$ functions, and so by
Theorem~\ref{t:edge-power} $f \in \clusp$.~\qed

\smallskip

We now show that if cluster machines 
with free boundaries no more powerful than the regular or circular
models then $\up = \coup$.
\begin{theorem}\label{t:free-circ-eq}
  The following are equivalent.
  \begin{enumerate}
  \item \label{t:free-circ-up=coup} $\up = \coup$.
\item \label{t:free-circ-reg} $\cluspfree = \clusp$.
\item \label{t:free-circ-circ} $\cluspfree \subseteq \cluspcircular$.
  \end{enumerate}
\end{theorem}
The technique used to prove Theorem~\ref{t:free-circ-eq} (one that we will 
later use to prove Theorem~\ref{t:p=up}) is, roughly speaking, to 
impose over an 
unambiguous or ``nearly unambiguous'' NPTM  an order that skips over any 
accepting
paths but then, at the end, sticks the skipped paths to the top of the order. Because
the free model imposes no restrictions on the last element, this is
relatively easy to do. But this makes accepting paths relatively
easy to locate in the other, ``non-free'' models, and that will allow
us to get $\coup \subseteq \up$.

\sproofof{\ref{t:free-circ-eq}} By Theorem~\ref{t:model-collapse}, 
\ref{t:free-circ-up=coup} 
$\Rightarrow$~\ref{t:free-circ-reg}. Because $\clusp \subseteq 
\cluspcircular$,~\ref{t:free-circ-reg} $\Rightarrow$~\ref{t:free-circ-circ}. 
For~\ref{t:free-circ-circ} $\Rightarrow$~\ref{t:free-circ-up=coup},
let $L \in \coup$. For some polynomial $p$, let $U'$ be a $p$-balanced NPTM 
such that $L(U') = \overline{L}$ and that, on any input, has at most one 
accepting path.
We may assume without loss of generality that $p > 0$ and that,
on any $x\in\Sigma^*$, $U'$ never accepts on path $1^{\pox}$. Let $U$ be
an NPTM that is the same as $U'$ except that, on any $x \in \Sigma^*$, $U$ on
input $x$ always accepts on path $1^{\pox}$. Thus $x \in L \Rightarrow U$ on
input $x$ has
exactly one accepting path and $x \not\in L \Rightarrow U$ on input $x$
has exactly two accepting paths. Define predicate $\uprec$ as follows. For 
$x,y,z \in \Sigma^*$, $\uprec(x,y,z)$ evaluates
to true exactly if $|y| = |z| = p(|x|)$ and:
\begin{enumerate}
\item $\{y,z\} \cap \acc_{U'}(x) = \emptyset$ and either
  \begin{enumerate}
  \item $y \lexprec z$, or
\item $(\exists w \in \acc_{U'}(x))[y \lexprec w \lexprec z]$,
  \end{enumerate}
or
\item $z\neq y=1^{\pox}$ and $z \in \acc_{U'}(x)$.
\end{enumerate}
Thus, $p$, $U$ and $\prec$ capture $\numacc_U$ in the sense of 
Definition~\ref{d:cluspfree}. 
(Note, however, that in defining $\prec$, which sets the adjacencies in
  $U$, we very deliberately refer to accepting paths of $U'$ rather
  than $U$, since this somewhat simplifies the definition of $\prec$.)

Suppose that $\numacc_U \in \cluspcircular$ via
$p'$, $M'$, and $\uprec'$. Then the NPTM $N$, 
described below, witnesses that $L \in \up$. On input $x\in \Sigma^*$,
$N$ nondeterministically guesses $y,z,w \in \Sigma^*$ where 
$|y|=|z|=|w|=p'(|x|)$ and accepts (on the current path)
exactly if $z \in \acc_{M'}(x)$,
$\{y,w\}\cap\acc_{M'}(x) = \emptyset$, $\uprec'(x,y,z)$, and 
$\uprec'(x,z,w)$.~\qed

\smallskip

The most pressing open question posed by
Theorems~\ref{t:model-collapse} and~\ref{t:free-circ-eq} and indeed by this 
paper is whether
$\up=\coup$ is a necessary condition for
$\clusp=\cluspcircular$.

Finally, we present three results that show that the free and circular
classes are in some ways relatively close to $\clusp$.

Let $\zof$ denote all 0-1-valued total functions, i.e., total
functions $f$ mapping from $\sigmastar$ to $\{0,1\}$.

\begin{theorem}\label{t:zero-one}
  $\cluspfree \cap \zof = \clusp \cap \zof$.
\end{theorem}

\sproof Let $f \in \cluspfree \cap \zof$ via some NPTM $M$ and
nondeterminism polynomial $p$. Then, for
all $x \in \Sigma^*$, $M$ has, at most, one accepting path. $M$ and $p$
together with $\prec$, $\bottom$, and $\top$ defined below witness that 
$f \in \clusp \cap \zof$.
For all $x\in \Sigma^*$, define $\bottom(x)$ as $0^{\pox}$ and $\top(x)$ as
$1^{\pox}$. For all $x, y, z \in \Sigma^*$, $\uprec(x,y,z)$ evaluates
to true exactly if $|y| = |z| = \pox$ and $y \lexprec z$.~\qed

\smallskip

Whether Theorem~\ref{t:zero-one} holds for 0-1-2-valued functions is
open.  (If we knew that the accepting-path cluster could without loss
of generality be assumed never to extend to the top or bottom element,
then 0-1-2-valued functions, $\bigo(1)$-valued functions, and much
more would work in Theorem~\ref{t:zero-one}, via
Theorem~\ref{t:edge-power}.  However, the desired ``without loss of
generality'' is not currently known to hold.)

Somewhat related to Theorem~\ref{t:zero-one} is the following result,
which shows that the sole obstacle to achieving
$\clusp=\cluspcircular$ is the possibility that the $\cluspcircular$
machine (i.e., $M$ of $p$, $M$, $\uprec$) has all paths accept on a
hard-to-predict set of inputs.  In particular, define
$\cluspcirculartilde$ to be the class of all $\cluspcircular$
functions $f$ whose membership in $\cluspcircular$ is instantiated by
some $p$, $M$, and $\uprec$ (in the sense of
Definition~\ref{d:circular}) such that
$\{x\condition f(x) = 2^\pox\}\in\p$. 

\begin{theorem}\label{t:tilde-tweak}
  $\cluspcirculartilde = \clusp$.
\end{theorem}

\sproof The $\subseteq$ direction follows easily from 
Theorem~\ref{t:edge-power}, since by
hypothesis we can test in $\p$ whether $f(x) = 2^\pox$ and with the
\emph{correct} answer in hand can produce in each case in a $\upsvp$
fashion
%
the accepting-block edge functions $\least$ and $\greatest$.
For the $\supseteq$ direction, let $p$, $M$, $\prec$, $\top$, and $\bottom$
capture $f$ in the sense of Definition~\ref{d:clusp}. Clearly,
$\{x\condition f(x) = 2^\pox\}\in\p$, via a deterministic polynomial-time
Turing machine that simulates $\top$ and $\bottom$. Let $\prec'$ be
the same as $\prec$, except that, for any $x \in \Sigma^*$,
$\uprec'(x,\top(x), \bottom(x))$ evaluates to true. Then
$p$, $M$, and $\prec'$ are a $\cluspcirculartilde$ model for $f$.~\qed

\smallskip

Considering Theorem~\ref{t:edge-power} and the above 
proof closely, it is not hard to 
see that one can prove the $\subseteq$ direction above even if 
in the definition of $\cluspcirculartilde$
one were to replace 
``$\{x\condition f(x) = 2^\pox\}\in\p$''
with ``$\{x\condition f(x) = 2^\pox\}\in \up \cap \coup$.''
And the $\supseteq$ direction of course holds with this adjustment.
Thus, Theorem~\ref{t:tilde-tweak} remains true even under that less
restrictive alternate definition.

Though Theorem~\ref{t:tilde-tweak} shows that $\cluspcircular$ and $\clusp$ have only one obstacle blocking their equality, the following result provides a bit of evidence against their equality. It shows that from equality there follows a consequence that we do not see how to establish without the assumption of equality. (We say that a function $g:\sigmastar\rightarrow\naturalnumber$ is \emph{strictly positive} if $(\forall x\in\sigmastar)[g(x)>0]$.)

\begin{theorem}\label{t:circ_eq_regular_consequence}
If $\clusp=\cluspcircular$, then for every $\clusp$ function $f$ there is a strictly positive $\fpt$ function $g$ such that $f+g$ is in $\clusp$.
\end{theorem}

The proof of this theorem relies on the following 
``complementarity'' property of the circular model.

\begin{proposition}\label{p:complementary}
Let $f$ be computed by $M$, $p$, $\prec$ in the circular model. Then the 
function $\overline{f}(x) = 2^{p(|x|)}-f(x)$ 
can be computed in the circular model.
\end{proposition}

\sproof  Let $M'$ be the Turing machine that is exactly the same as $M$ except
that, for each $x \in\Sigma^*$ and each path $\rho$ of $M(x)$, 
if $M(x)$ halts and rejects on path $\rho$ then $M'(x)$ halts
and accepts on its analogous path,
and if $M(x)$ halts and accepts on path $\rho$
then $M'(x)$ halts and
rejects on its analogous path. 
We claim that $M'$, $p$, and $\prec$ compute $\overline{f}$ in the circular
model. To see this,
choose $x\in\sigmastar$. Suppose that 
$0<f(x)<2^{p(|x|)}$. Recall, as per the paragraph
immediately following Definition~\ref{d:circular},
that we are allowed to adjust the bijection
$h_x$ in such a way that $h_x^{-1}(0^{p(|x|)})$ is an accepting 
path
of $M(x)$ and $h_x^{-1}(1^{p(|x|)})$ is a rejecting path.
Thus, we may assume that the first element under the image of $h_x$ is
the least accepting path 
of $M$. Then, obviously, the last element under the image
of $h_x$ is the greatest accepting path of $M'$ on input $x$.
If $f(x) \in \{0,2^{p(|x|)}\}$ then $h_x$
witnessing that the computational paths of $M$ on input $x$ are
clustered in the circular model will also witness that the computational paths 
of $M'$ on input $x$ are clustered in the circular model. Since
we do not change $p$ or $\prec$ (or $h_x$), we easily see that
$M'$ computes exactly $\overline{f}(x)=2^{p(|x|)}-f(x)$ in
the circular model.~\qed

\smallskip

\sproofof{Theorem \ref{t:circ_eq_regular_consequence}} Let $f$ be a
$\clusp$ function via NPTM $M$, nondeterminism polynomial $p$,
predicate $\prec$, $\fpt$ functions $b$ and $t$. Without loss of
generality, we may assume that for all $x\in\Sigma^*$, $f(x)<
2^{p(|x|)}$.\footnote{Otherwise consider an NPTM $N$ that on input
  $x\in\Sigma^*$, guesses a string $u\in\{0,1\}^{p(|x|)+1}$, simulates
  $M$ along path $\suffix(u,p(|x|))$ if the first bit of $u$ is zero, and
  otherwise rejects along the path $u$. For all $x,y,z\in\Sigma^*$,
  $a,b\in\Sigma$, the corresponding predicate $\uprec'(x,ay,bz)$
  evaluates to true if and only if $[a=b \land \uprec(x,y,z)]$ or $[a=0
  \land y=t(x)\land b=1\land z=b(x)]$. Bottom and top functions $b'$
  and $t'$ are given by $b'(x)=0b(x)$ and $t'(x)=1t(x)$.}  Clearly,
$f$ is in $\cluspcircular$ via the same NPTM $M$, the same
nondeterminism polynomial $p$, and the new predicate $\prec'$ that is
the same as $\prec$ except that, for any $x\in\Sigma^*$,
$\uprec'(x,\top(x),\bottom(x))$ evaluates to true.  Thus, by
Proposition \ref{p:complementary} the function $\overline{f}$, defined
for all $x\in\Sigma^*$ as $\overline{f}(x)=2^{p(|x|)}-f(x)$, belongs
to $\cluspcircular$. Define the function $f'$ on each $x\in\Sigma^*$
as $f'(x)=\max\{0, \overline{f}(x)-1\}$ ($=\overline{f}(x)-1$, because
$\overline{f}$ is strictly positive). From the assumption
$\clusp=\cluspcircular$ and Theorem \ref{t:positive_closures}, we
conclude that $f'\in\clusp$, say, using the nondeterminism polynomial
$q$. Here, we may assume that for all $x\in\Sigma^*$, $q(|x|)\ge
p(|x|)$.  
Again, we have $f'\in\cluspcircular$ and this can be
witnessed using the same polynomial $q$. Now, let $\overline{f'}$ be
the function defined for all $x\in\Sigma^*$ as
$\overline{f'}(x)=2^{q(|x|)}-f'(x)$. Then, by Proposition
\ref{p:complementary}, $\overline{f'}$ is a $\cluspcircular$ function
and thus, once more by our assumption, a $\clusp$ function. Define an
$\fpt$ function $g$, for each $x\in\Sigma^*$, as
$g(x)=2^{q(|x|)}-2^{p(|x|)}+1$. It follows that for all
$x\in\Sigma^*$, $g(x)\ge 1$ and $\overline{f'}(x)=2^{q(|x|)}-
(2^{p(|x|)}-f(x)-1) = f(x) + g(x)$. So $f+g\in\clusp$.~\qed

\smallskip

Finally, we have the following result, which shows that if it does
hold that $\clusp=\cluspfree$ or $\clusp=\cluspcircular$, then we
probably can expect that $\clusp$ will at least need to in some cases
use more nondeterminism than the other two classes.

\begin{theorem}~\label{t:p=up}
\begin{enumerate}
  
\item\label{it:circ-pup} If for each $p$, $M$, and $\uprec$ that instantiate a
  $\cluspcircular$ function that function is also instantiated by a
  $\clusp$ machine having nondeterminism exactly $p$, then $\p=\up$.
  
\item\label{it:free-pup}If for each $p$, $M$, and $\uprec$ that instantiate a
  $\cluspfree$ function that function is also instantiated by a
  $\clusp$ machine having nondeterminism exactly $p$, then $\p=\up$.

\item\label{it:upsvt-pup} If each $\cluspfree$ machine has $\upsvt$ functions $\top$ and $\bottom$,
 then $\up = \coup$.
\item\label{it:fpt-pup} If each $\cluspfree$ machine has $\fpt$ functions $\top$ and $\bottom$,
then $\p = \up$.
\end{enumerate}
\end{theorem}

\sproof 
Let $L \in \up$ and let $U$ be an 
NPTM such that $L(U) = L$ and that $U$ has on any
input at most one accepting path. Assume without loss of generality 
that $U$ is balanced via polynomial $p$ where $p>0$ and that $U$ never accepts 
on path
$1^{p(|x|)}$. Let $U'$ be an NPTM that is the same 
as $U$ except that the accepting and rejecting states are switched.

For item~\ref{it:circ-pup}, define $\prec''$ as follows: For all $x,y,z \in 
\Sigma^*$, $\uprec''(x,y,z)$ evaluates to true exactly if $|y| = |z|$ and either
$y \lexprec z$ or ($y \in 1^*$ and $z \in 0^*$). Clearly,
$U'$, $p$, and $\uprec''$ captures $\numacc_{U'}$ in the sense of 
Definition~\ref{d:circular}.

For items~\ref{it:free-pup},~\ref{it:upsvt-pup}, and~\ref{it:fpt-pup}, define
 $\prec$ as follows: For all 
$x, y,z \in \Sigma^*$, $\uprec(x,y,z)$ evaluates to true exactly if:
\begin{enumerate}
\item $\{y,z\} \subseteq \acc_{U'}(x)$ and either
  \begin{enumerate}
  \item $y \lexprec z$, or
\item $(\exists w \not\in \acc_{U'}(x))[y\lexprec w \lexprec z]$,
  \end{enumerate}
or
\item $y=1^{\pox}$ and $z\not\in\acc_{U'}(x)$.
\end{enumerate}
Clearly, $p$, $U'$, and $\uprec$ captures $\numacc_{U'}$ in the sense of 
Definition~\ref{d:cluspfree}.

Since we are only considering cluster machines, all
the accepting paths must be contiguous with respect to $\prec$, and this
forces any rejecting path to occur as the bottom or the top path. Under
the assumptions of item~\ref{it:upsvt-pup}, then, $\overline{L} \in \up$
by a machine that simulates in sequence $\upsvt$ machines for the $\top$
and $\bottom$ functions of $p$, $U'$, and $\uprec$. Under the assumptions
of item~\ref{it:fpt-pup}, $L \in \p$ via a deterministic polynomial-time
Turing machine that simulates in sequence the $\top$ and $\bottom$ functions
of $p$, $U'$, and $\uprec$. 

For items~\ref{it:free-pup} and ~\ref{it:circ-pup}, suppose
that $\numacc_{U'}\in\clsp$ via some $M$, $p'$, $\bottom'$, $\top'$, $\prec'$
and suppose that 
$p = p'$. Then $L \in P$ via a deterministic, polynomial-time
Turing machine that, on input $x \in \Sigma^*$ accepts if and only if
$\{\bottom'(x),\top'(x)\} \not\subseteq \acc_M(x)$. Note that because $M$
has the same amount of nondeterminism as $U'$ it follows that $M$ has, at most,
one rejecting path (which must occur at the top or bottom of the order) and 
this path occurs if and only if $x \in L$.~\qed



\paragraph*{Acknowledgments} We are deeply grateful to Klaus W. Wagner
for hosting the visit to W\"urzburg during which much of this work was
done, for proposing the class $\cluspfree$, and for many other
valuable suggestions and insights.


\bibliography{gry}

\newcommand{\etalchar}[1]{$^{#1}$}
\begin{thebibliography}{HHKW05}

\bibitem[BDG90]{bal-dia-gab:b:sctII}
J.~Balc{\'{a}}zar, J.~D{\'{\i}}az, and J.~Gabarr{\'{o}}.
\newblock {\em Structural Complexity {II}}.
\newblock EATCS Monographs in Theoretical Computer Science. Springer-Verlag,
  1990.

\bibitem[BDG95]{bal-dia-gab:b:sctI-2nd-ed}
J.~Balc{\'{a}}zar, J.~D{\'{\i}}az, and J.~Gabarr{\'{o}}.
\newblock {\em Structural Complexity {I}}.
\newblock EATCS Texts in Theoretical Computer Science. Springer-Verlag, 2nd
  edition, 1995.

\bibitem[BLS84]{boo-lon-sel:j:quant}
R.~Book, T.~Long, and A.~Selman.
\newblock Quantitative relativizations of complexity classes.
\newblock {\em SIAM Journal on Computing}, 13(3):461--487, 1984.

\bibitem[BLS85]{boo-lon-sel:j:qual}
R.~Book, T.~Long, and A.~Selman.
\newblock Qualitative relativizations of complexity classes.
\newblock {\em Journal of Computer and System Sciences}, 30(3):395--413, 1985.

\bibitem[CGH{\etalchar{+}}88]{cai-gun-har-hem-sew-wag-wec:j:bh1}
J.~Cai, T.~Gundermann, J.~Hartmanis, L.~Hemachandra, V.~Sewelson, K.~Wagner,
  and G.~Wechsung.
\newblock The boolean hier\-archy {I}: {S}truc\-tural proper\-ties.
\newblock {\em SIAM Jour\-nal on Com\-pu\-ting}, 17(6):1232--1252, 1988.

\bibitem[GH00]{gla-hem:j:clarityI}
C.~Gla{\ss}er and L.~Hemaspaandra.
\newblock A moment of perfect clarity {I}: {T}he parallel census technique.
\newblock {\em SIGACT News}, 31(3):37--42, 2000.

\bibitem[GS88]{gro-sel:j:complexity-measures}
J.~Grollmann and A.~Selman.
\newblock Complexity measures for public-key cryptosystems.
\newblock {\em SIAM Journal on Computing}, 17(2):309--335, 1988.

\bibitem[HHKW05]{hem-hom-kos-wag:tr-special-with-backpointer-for-cluster-new-p%
aper-to-icalp-version:interval-functions}
L.~Hemaspaandra, C.~Homan, S.~Kosub, and K.~Wagner.
\newblock The complexity of computing the size of an interval.
\newblock Technical Report TR-856, Department of Computer Science, University
  of Rochester, Rochester, NY, February 2005.
\newblock Revised, March 2005. This is an expanded version
  of~\protect\cite{hem-kos-wag:cOutByhem-hom-kos-wagTR:interval-functions}.

\bibitem[HKW01]{hem-kos-wag:cOutByhem-hom-kos-wagTR:interval-functions}
L.~Hemaspaandra, S.~Kosub, and K.~Wagner.
\newblock The complexity of computing the size of an interval.
\newblock In {\em Proceedings of the 28th International Colloquium on Automata,
  Languages, and Programming}, pages 1040--1051. Springer-Verlag {\it Lecture
  Notes in Computer Science \#2076}, July 2001.

\bibitem[Kos99]{kos:j:clusters}
S.~Kosub.
\newblock A note on unambiguous function classes.
\newblock {\em Information Processing Letters}, 72(5--6):197--203, 1999.

\bibitem[MP79]{mey-pat:t:int}
A.~Meyer and M.~Paterson.
\newblock With what frequency are apparently intractable problems difficult?
\newblock Technical Report MIT/LCS/TM-126, Laboratory for Computer Science,
  MIT, Cambridge, MA, 1979.

\bibitem[Sch76]{sch:c:self-reducible}
C.~Schnorr.
\newblock Optimal algorithms for self-reducible problems.
\newblock In {\em Proceedings of the 3rd International Colloquium on Automata,
  Languages, and Programming}, pages 322--337. Edinburgh University Press, July
  1976.

\bibitem[Sel90]{sel:t:adaptive}
A.~Selman.
\newblock A note on adaptive vs.\ nonadaptive reductions to {N}{P}.
\newblock Technical Report 90-20, Department of Computer Science, State
  University of New York at Buffalo, Buffalo, NY, September 1990.

\bibitem[Sel94]{sel:j:taxonomy}
A.~Selman.
\newblock A taxonomy of complexity classes of functions.
\newblock {\em Journal of Computer and System Sciences}, 48(2):357--381, 1994.

\bibitem[VW95]{vol-wag:j:opt}
H.~Vollmer and K.~Wagner.
\newblock Complexity classes of optimization functions\typeout{MINOR PANIC: get
  year, number, pages,etc}.
\newblock {\em Information and Computation}, 120(2):198--219, 1995.

\bibitem[Wag90]{wag:j:bounded}
K.~Wagner.
\newblock Bounded query classes.
\newblock {\em SIAM Journal on Computing}, 19(5):833--846, 1990.

\end{thebibliography}


\typeout{ REMINDER TO OURSELVES: sigh...  To avoid an ambiguity we
  have made most function names italic especially t for top...  but
  total and partial functions have a t/p Roman subscript.  Be VERY
  careful regarding galleys eventually, as copyeditors will ruin this,
  possibly.  *END OF REMINDER TO OURSELVES* *END OF REMINDER TO
  OURSELVES* *END OF REMINDER TO OURSELVES* *END OF REMINDER TO
  OURSELVES* *END OF REMINDER TO OURSELVES* *END OF REMINDER TO
  OURSELVES* }

\end{document}